\documentclass[showpacs, pra,onecolumn,preprintnumbers ,amsmath, amssymb, superscriptaddress, aps]{revtex4-2}
\usepackage{color}
\usepackage{amsmath,amssymb}
\usepackage{pifont}
\usepackage{amssymb}  
\usepackage{bbold}
\usepackage{float}
\usepackage{subfloat}

\usepackage[caption=false]{subfig}
\usepackage{tikz}
\usepackage{makecell}
\usepackage{subfig}
\usepackage{pifont}   
\usepackage{graphicx} 
\graphicspath{{Figures/}}
\usepackage{dcolumn}  
\usepackage{bm}       
\usepackage{multirow} 
\usepackage{placeins}
\usepackage[colorlinks]{hyperref}
\usepackage{mathtools}
\usepackage{appendix}

\begin{document}
	\title{Virtual  excitations and quantum correlations in ultra-strongly coupled  harmonic oscillators under intrinsic decoherence }
	\date{\today}
	\author{Radouan Hab-arrih}
	\email{habarrih46@gmail.com}
	\affiliation{Laboratory of Theoretical Physics, Faculty of Sciences, Choua\"ib Doukkali University, PO Box 20, 24000 El Jadida, Morocco}
	\author{Ahmed Jellal}
	\email{a.jellal@ucd.ac.ma}
	\affiliation{Laboratory of Theoretical Physics, Faculty of Sciences, Choua\"ib Doukkali University, PO Box 20, 24000 El Jadida, Morocco}
	\affiliation{Canadian Quantum  Research Center,
				204-3002 32 Ave Vernon,  BC V1T 2L7,  Canada}
	\author{El Hassan El Kinani}
	\affiliation{Mathematical Modeling and Scientific Computing Team, Department of Mathematics, Faculty of Sciences, Moulay Ismail University, Meknes, Morocco}
	\pacs{ 
	}
	\begin{abstract}
We study the intrinsic decoherence of coupled harmonic oscillators. The Milburn master equation is solved exactly, and the dynamics of virtual ground state excitations are investigated.  The interaction of quantum correlations and virtual excitation  was then studied. 
The following is a summary of our major findings.
(i) The damped oscillatory profile of all three quantities is the same.
(ii) Ultra-strong coupling combined with huge anisotropy values results in the reemergence of entanglement and steering.
(iii) To sustain entanglement and steering, virtual excitations are required. (iv)
The quantum correlations are amplified in the quantum synchronous regime.
(v) Ultra-strong couplings cause inherent decoherence to be avoided. 
\end{abstract}

\maketitle

	\section{Introduction}
Entanglement is one of the amazing resource of quantum mechanics that  have not  any analog in classical arena.  In quantum information processing, such resource must be preserved during all scenarios. Unfortunately, because of the  coherence losses (decoherence), entanglement dies out rapidly in finite time, such phenomenon is called entanglement sudden death (ESD) \cite{ESD1}. Recently, it was shown by engineering the physical parameters  entanglement revives after its ESD, the issue that brought some hope  on quantum computing and the promised quantum computer\cite{R03,R04}. 
In 1935, Einstein, Podolsky and Rosen (EPR), proposed an intellectual experience to refute quantum entanglement for the completeness of quantum theory (QT) \cite{EPR}. Later on, Schr\"{o}dinger introduced another quantum correlation, the so-called quantum steering to support QT \cite{ster}. In the modern quantum information science (QIS), the quantum steering 
denotes a quantum correlation situated between entanglement and Bell non locality \cite{ster1}. It denotes 
the impossibility to describe the conditional states at one
party by a local hidden state model \cite{ster3}. Moreover, recently, the steering is widely discussed in several systems, including continous variables \cite{hierarchy,magnon,symmetry} and discrete systems \cite{qubitsteering, qubitsteering2}.

Ultrastrong coupling (USC) physics plays a paramount role in several fields, ranging from quantum optics to condensed matter \cite{USC1,USC2}. USC requires that the coupling $ J $ between the system parts be of the same order of magnitude of the transition frequencies $ \omega $ of the system $ (0.1<\frac{J}{\omega^{2}}<1) $. Furthermore, the USC regime has achieved important records in several experimental setting, let's quote, for example, intersubband polaritons \cite{polariton}, superconducting circuits \cite{superconducting}, Landau polaritons \cite{LP}  and optomechanics \cite{optomechanic}. The importance of  USC regime lies in its drastic effects on some standard fundamental physical effects, including, for instance, the Purcell effect \cite{Purcell}, Zeno effect \cite{Zeno}, photon blockade \cite{PB} and excitations in the ground state \cite{maintaining}.   

The ground state excitations are one of the most amazing predictions of modern quantum theory. In USC, the ground state is not empty, but
populated with a sea of virtual particles. These short-lived 
fluctuations are originating from the couter-rotating terms of the Hamiltonian, because in USC the wave rotating approximation (WRA) in not valid  and anti-resonant terms should be kept in the Hamiltonian \cite{WRA,WRa}. Those excitations are the background of some of the most important
physical processes in the universe, namely, the Casimir effect \cite{Casimir} and Hawking radiation \cite{Hawking}.
The main open theoretical challenge in USC is to distinguish between virtual (unobservable) and physical (observable) excitations \cite{virtual}. Recently, ground state excitations have been extensively studied, and it has been demonstrated that they cannot be detected \cite{DET1,DET2}. However, some authors have shown the possibility of indirect (without being absorbed) detection, for instance, by measuring the change that they produce in the Lamb shift of an ancillary probe qubit coupled to the cavity \cite{Lamb}, or by probing the radiation pressure that they generate in optomechanics systems \cite{pressure}.
Although these excitations are not directly detectable, they have the potential to spontaneously change into real excitations \cite{DET2}. 

 Intrinsic decoherence (ID) is a mechanism by which the system loses its coherence because of its intrinsic degrees of freedom. Several models of ID have been proposed \cite{ID1,ID2,ID3}, and one of them, is Milburn decoherence (MD) model. 
 Milburn proposed a simple modification to the Schr\"odinger equation based on the assumption that the change in the state of the system, i.e., $\rho(t)\rightarrow \rho(t+\tau)$, is uncertain and occurs with a probability of $ 0<p(\tau)<1 $   on a sufficiently small time scale \cite{Milburn}. The change of state is always certain in standard quantum mechanic, and $ p(\tau)=1 $.
%
 In the last decade, the ID of quantum coupled  discrete systems has been widely studied  \cite{D1,D2,D3,D4}, but the majority of those works dealt with diffusion approximation, i.e., first order correction of the von Neumann equation. 
 
 One of the important continuous systems is the quantum-coupled harmonic oscillator (QCHO). It models, for example, coupled ions in ion traps \cite{trap}, coupled nano-sized electromechanical devices arrayed \cite{Nano}, and light propagation in inhomogeneous media \cite{light}. 
%
  In the last decades, the environmental decoherence of QCHO has been well studied in both regimes of coupling weak and ultrastrong \cite{osci, maintaining}, as well as  Markovian and non-Markovian evolution \cite{markov, markov1}. Fortunately, it was shown that avoiding environmental decoherence is possible by engineering only the frequencies and coupling between oscillators \cite{avoiding}.  The effect of MD on  coupled oscillators beyond WRA  is not yet studied. In this paper we shed light on the effect of MD on the virtual excitations of the ground state and their interconnection with quantum entanglement and steering. Subsequently, we discuss how one can avoid intrinsic decoherence by appropriately choosing the experimentally accessible parameters.   
  
  The following is a breakdown of the current paper's structure.
  The Hamiltonian is diagonalized in the creation and annihilation operators in Sec. \ref{sec1}.
  We construct the covariance matrix and quantum correlations with virtual excitations in Sec. {\ref{sec3}.
  In Sec. \ref{sec4}, the numerical results and discussions will be presented.
  Finally, we present a quick summary of our findings
%

\section{Diagonalization and Milburn dynamics \label{sec1}}

Consider a system consisting of two coupled harmonic oscillators that can be described by the Hamiltonian (in the unit $ \hbar=m=1 $)  
%
\begin{eqnarray}
\hat{H}(\hat{x}_{1},\hat{x}_{2})= \frac{\hat{p}_{1}^{2}}{2}+\frac{\hat{p}_{2}^{2}}{2}+\frac{1}{2}\omega_{1}^{2}\hat{x}_{1}^{2}+%
\frac{1}{2}\omega_{2}^{2}\hat{x}_{2}^{2}-J \hat{x}_{1}\hat{x}_{2}
\end{eqnarray}
\begin{figure}[H]
	\centering
	\includegraphics[width=11cm, height=6cm]{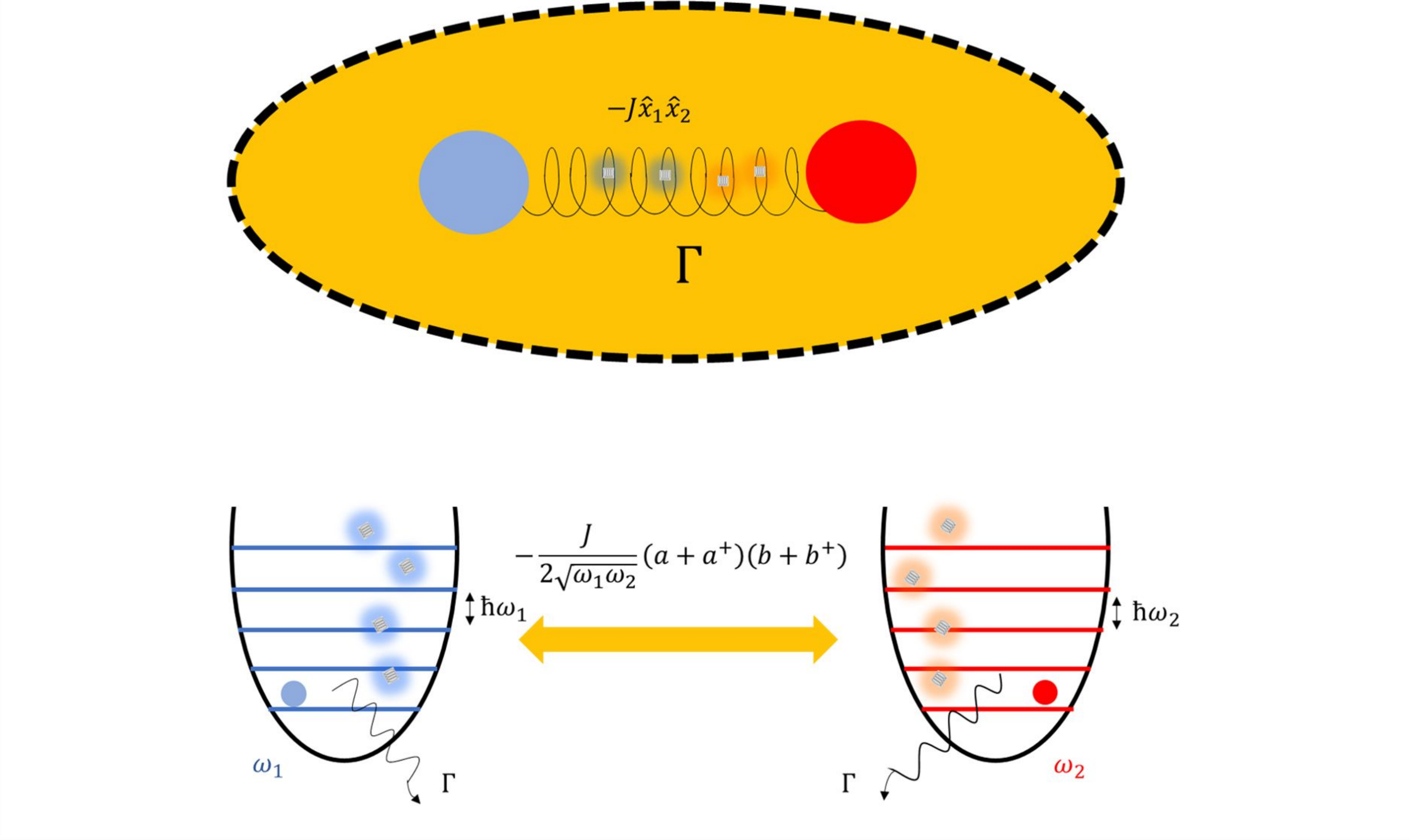}
	\caption{(color online) Two harmonic oscillators coupled via position-position type interaction $-J\hat{x}_{1}\hat{x}_{2}$, the system losses its coherence via Milburn intrinsic decoherence characterized by the factor $\Gamma$. The excitations in the ground state are virtual (not being absorbed) and the intrinsic decoherence affects their generation.  }
\end{figure}
To diagonalize the Hamiltonian, we perform a unitary transformation around the $ z $-axis
performing a rotation $\mathcal{R}(\theta)$ a round $ z- $axis, such that 
\begin{align}
\mathcal{R}(\theta)=\begin{pmatrix}
\cos\theta & -\sin\theta \\ 
\sin\theta & \cos\theta
\end{pmatrix}  
\label{rot}
\end{align}
 of angle  given by
\begin{eqnarray}
\theta
= \frac{1}{2} \arctan\left(\frac{2g}{R^{4}-1}\right) 
\end{eqnarray}
where 
the dimensionless parameters $R=\sqrt{\omega_{1}/\omega_{2}}$ and 
$g=J/\omega_{2}^{2}$ have been defined. 
It is worth noting that when $\omega_{1}=\omega_{2}$, the rotation angle becomes $\theta= \frac{\pi}{4}$.
Now the  Hamiltonian takes the following form 
\begin{align}\label{ham4}
\hat{H}_{D}(\hat{x}_{1},\hat{x}_{2})=\frac{\hat{p}_{1}^{2}}{2}+\frac{1}{2}\Omega_{1}^{2}\hat{x}_{1}^{2}+\frac{\hat{p}_{2}^{2}}{2}+%
\frac{1}{2}\Omega_{2}^{2}\hat{x}_{2}^{2}
\end{align}
and the normal frequencies  are  (we assume that $ \omega_1>\omega_ 2 $ without loss of generality)
\begin{align}
\Omega_{1}^{2}/\omega_{1}^{2}&= (1+R^{-4})/2+\frac{1}{2}\sqrt{(1-R^{-4})^{2}+4g^{2}R^{-8}}\\
\Omega_{2}^{2}/\omega_{2}^{2}&= (1+R^{4})/2-\frac{1}{2}\sqrt{(1-R^{4})^{2}+4g^{2}}.
 \end{align} 
We introduce a pair of commuting operators
\begin{align}
	A=\tfrac{1}{2\sqrt{\Omega_{1}}}(\Omega_{1}\hat{x}_{1}+i\hat{p}_{1}), \qquad B=\tfrac{1}{2\sqrt{\Omega_{2}}}(\Omega_{2}\hat{x}_{2}+i\hat{p}_{2})
\end{align}
and then write  Eq. \eqref{ham4} as
\begin{align}
	\hat{H}_{D}(\hat{x}_{1},\hat{x}_{2})	
	=\Omega_{1}A^+A+\Omega_{2}B^{+}B+\frac{\Omega_{1}+\Omega_{2}}{2} \label{diagonal}.
\end{align}
The following quantum squeezers, expressed in terms the original  commuting operators $ a $ and $ b  $ 
\begin{eqnarray}
\mathcal{S}_{w}(s_{j})=\exp[s_{j}(w^{2}-w^{+ 2})], \qquad w=a,b
\label{squeeze}\end{eqnarray}
are used to diagonalize the Hamiltonian in the original creation and annihilation operators,  with
 $j=1,2$ and the squeezing parameters  $ s_{j}=1/2\ln(\Omega_{j}/\omega_{j}) $. As a result, we obtain 
\begin{eqnarray}
\hat{H}_{d}=\Omega_{1}a^+a+\Omega_{2}b^{+}b+\frac{\Omega_{1}+\Omega_{2}}{2}.
\end{eqnarray}

The Milburn master equation will be used to characterize the system's intrinsic decoherence \cite{Milburn}. This is 
\begin{equation}
\dot{\rho}(t)=\Gamma\left\lbrace  \exp \left[ -\frac{i}{\Gamma}\hat{H}\right] \rho(0) \exp \left[ \frac{i}{\Gamma}\hat{H}\right] -\rho \right\rbrace 
\end{equation}
where $\Gamma$ is the intrinsic damping factor. Its formal solution 
is given by \cite{formal,Moya2022}
\begin{eqnarray}
\rho(t,\Gamma)=\exp\left[ -\Gamma t \right] \sum_{k=0}^{\infty}\frac{(\Gamma t)^k}{k!}\exp\left[ -i\frac{k}{\Gamma} \hat{H}\right] \rho(0)\exp\left[ i\frac{k}{\Gamma} \hat{H}\right] \label{condition initiale}.
\end{eqnarray}
We note here that the decoherence is rather high for  $ \Gamma  \rightarrow 0 $ and extremely weak for 
  $ \Gamma \rightarrow +\infty $ \cite{Milburn}. 
  The goal of our work  is to investigate a fundamental feature of the ground state, namely virtual excitation. Then, consider that oscillators are initially in their vacuum state given by
%
\begin{eqnarray}
\rho(0)=|0\rangle_1\langle 0|\times |0\rangle_2\langle 0|=|0 0\rangle\langle 0 0|.
\end{eqnarray}
As a matter of convenience, we set 
\begin{align}
&|\Phi_{k}(\Gamma)\rangle= \exp\left[ -i\frac{k}{\Gamma} \hat{H}\right] |0 0\rangle\\
& 0\leq\varOmega_{k}(t,\Gamma)=\exp[-\Gamma t]\frac{(\Gamma t)^{k}}{k!}\leq 1.
\end{align}
Thus, Eq. (\ref{condition initiale}) becomes 
\begin{eqnarray}
\rho(t,\Gamma)= \sum_{k=0}^{\infty} \varOmega_{k}(t,\Gamma)  |\Phi_{k}(\Gamma)\rangle \langle \Phi_{k}(\Gamma)|\label{10}.
\end{eqnarray}
Finally, the time evolution of the observable $\mathcal{O}$ is 
\begin{eqnarray}
\langle{\mathcal{O}}\rangle(t)= \sum\limits_{k=0}^{+\infty}\varOmega_{k}(t,\Gamma)\langle \Phi_{k}(\Gamma)|\mathcal{O}|\Phi_{k}(\Gamma)\rangle.\label{16}
\end{eqnarray}

\section{Quantum correlations and virtual excitations \label{sec3} }
\subsection{Covariance formalism}
In our case, we consider two bosonic modes, $a$ and $b$, which satisfy the bosonic algebra commutator $[\mathfrak{o},\mathfrak{o}^{+}]=\mathbb{I} $ $(\mathfrak{o}=a,b)$. We collect these operators in the vector $ \mathbb{A}=(\mathbb{A}_{n})=(a,a^{+},b,b^{+})^{T}$, while $T$ stands for the transpose. 
The above commutation relation between the creation and annihilation operators can be rewritten in the matrix form 
$[\mathbb{A}_{n},\mathbb{A}_{m}]=i\mathbb{J}_{n,m}$. It is simple to write $\mathbb{J}$ in the symplectic form shown below
\begin{eqnarray}
i\mathbb{J}_{4\times 4}&=&\left(\begin{array}{cccc}
   \tilde{\mathbb{{I}}}_{2\times 2}  &  0\\
   0  & \tilde{\mathbb{{I}}}_{2\times 2}
\end{array}\right), \qquad \tilde{\mathbb{{I}}}_{2\times2}=\left(\begin{array}{cc}
     0&  1\\
     -1& 0
\end{array}\right).
\end{eqnarray}
A symplectic matrix $\mathbb{S}$ is one that verifies the following properties: $\mathbb{S}^{+}\mathbb{J}\mathbb{S}=\mathbb{J}$ and $\det(\mathbb{S})=1$. It is worth noting that the corresponding symplectic matrix for a given unitary operator $\mathbb{U}$ can be derived using the fact that $\mathbb{U}\mathbb{A}\mathbb{U}^{+}=\mathbb{S} \mathbb{A}$. To quantify quantum correlations together with virtual excitations, we use the covariance matrix $\Sigma$ defined by the compact form
\begin{eqnarray}
\Sigma_{n,m}&=&\langle \lbrace \mathbb{A}_{n},\mathbb{A}^{+}_{m}
\rbrace\rangle_{\rho}-2\langle \mathbb{A}_{n}\rangle_{\rho}\langle\mathbb{A}^{+}_{m}\rangle_{\rho}.
\end{eqnarray}
We mention that when $ \langle \mathbb{A}_{n}\rangle_{\rho}=\langle\mathbb{A}^{+}_{m}\rangle_{\rho}=0$, the covariance matrix reduces to $\Sigma_{n,m}=2\langle\mathbb{A}_{n}^{+}\mathbb{A}_{m}\rangle+1$. Here, all the averages will be computed by using Eq. (\ref{16}), and $\lbrace x,y\rbrace:=xy+yx$. It is known that every quantifier of entanglement
for two mode symmetric Gaussian states is a function of
the smallest symplectic eigenvalue of the partial transpose \cite{symplectic}.

\subsection{Symplectic form of unitary transformations}

In the annihilation and creation representation, the rotation unitary operator Eq. (\ref{rot}) can be reduced to 
\begin{eqnarray}
\mathcal{R}(\theta)= \exp\left[\tfrac{\theta}{2}\tfrac{1-R^2}{R}(a^+b^+-ab)- \tfrac{\theta}{2}\tfrac{1+R^2}{R}(a^+b-ab^+)\right]. 
\end{eqnarray}
We show the results
\begin{align}
\mathcal{R}(\theta) a \mathcal{R}^{-1}(\theta)&= a \cos\theta+\tfrac{1}{2}b\sin(\theta)(R+R^{-1})+\tfrac{1}{2}b^{+}\sin (\theta) (R-R^{-1}) \label{23}\\
\mathcal{R}(\theta) b \mathcal{R}^{-1}(\theta)&= b \cos\theta-\tfrac{1}{2}a\sin(\theta)(R+R^{-1})+\tfrac{1}{2}a^{+}\sin (\theta) (R-R^{-1})\label{24}. 
\end{align}
As a result,  the symplectic form $\mathbb{S}_{\mathcal{R}}$ of $\mathcal{R}(\theta)$ is obtained
\begin{eqnarray}
\mathbb{S}_{\mathcal{R}}(\theta)=\begin{pmatrix}
   \cos\theta &  0&\frac{R+R^{-1}}{2}\sin\theta&\frac{R-R^{-1}}{2}\sin\theta\\
0  & \cos\theta&\frac{R-R^{-1}}{2}\sin\theta&\frac{R+R^{-1}}{2}\sin\theta\\
      -\frac{R+R^{-1}}{2}\sin\theta&\frac{R-R^{-1}}{2}\sin\theta&\cos\theta&0\\
   \frac{R-R^{-1}}{2}\sin\theta&-\frac{R+R^{-1}}{2}\sin\theta&0&\cos\theta\\
\end{pmatrix}.\label{rot}
\end{eqnarray}
Subsequently, one can rewrite the squeezors defined in Eq. (\ref{squeeze}) in their symplectic form
\begin{eqnarray}
\mathbb{S}_ {\mathcal{S}}(s_{1},s_{2})=
\begin{pmatrix}
     \cosh s_1&\sinh s_1&0&0  \\
     \sinh s_1&\cosh s_1&0&0\\
 0&0&\cosh s_2&\sinh s_2\\
     0&0&\sinh s_2 &\cosh s_2
 \end{pmatrix}.\label{squeez}
\end{eqnarray}
Thus, we show that the symplectic transformation corresponding to
$\exp\left[-i\frac{k}{\Gamma}H_{d}\right]$
takes the following form
\begin{eqnarray}
\mathbb{S}_{H}(\Gamma)= \text{diag}\left(\exp\left[-i\frac{k}{\Gamma}\Omega_{1}\right],\exp\left[i\frac{k}{\Gamma}\Omega_{1}\right],\exp\left[-i\frac{k}{\Gamma}\Omega_{2}\right], \exp\left[i\frac{k}{\Gamma}\Omega_{2}\right]\right)\label{diag}
\end{eqnarray}
\subsection{Virtual excitations and quantum correlations}
To compute virtual excitations and quantum correlations encoded in our state, we use the covariance matrix formalism. It is not difficult to show that the covariance matrix $\Sigma$ 
is
\begin{eqnarray}
\Sigma(t,\Gamma)= \exp[-\Gamma t]\sum\limits_{k=0}^{\infty}\frac{(\Gamma t)^k}{k!}
 \mathcal{H}(\theta,\Gamma,R) \Sigma(0)\mathcal{H}^{\dagger}(\theta,\Gamma,R)\label{haM}
\end{eqnarray} 
where  the symplectic matrix $\mathcal{H}$ is defined as 
\begin{eqnarray}
\mathcal{H}= \mathbb{S}_{\mathcal{R}}(-\theta, R)\mathbb{S}_{\mathcal{S}}(s_1,s_2)\mathbb{S}_{H}(\Gamma)\mathbb{S}_{\mathcal{S}}(-s_1,-s_2)\mathbb{S}_{\mathcal{R}}(\theta, R)
\end{eqnarray}
and $\Sigma(0)$ is the covariance matrix corresponding to vacuum state $\rho(t=0)=|00\rangle\langle 00|$
\begin{eqnarray}
\Sigma(0)= \begin{pmatrix}
     1&0&0&0 \\
     0&1&0&0\\
     0&0&1&0\\
     0&0&0&1
\end{pmatrix}
\end{eqnarray}
with $x^{\dagger}$ denotes the hermitian conjugate of $x$.
After computation, we end up with 
\begin{eqnarray}
 \Sigma(t,\Gamma)=\begin{pmatrix}
     \sigma_{11}&\sigma_{12}&\sigma_{13}&\sigma_{23}^{\ast} \\
     \sigma_{12}^{\ast}&\sigma_{11}&\sigma_{23}&\sigma_{13}^{\ast}\\
     \sigma_{13}^{\ast}
     &\sigma_{23}^{\ast}&\sigma_{33}&\sigma_{34}\\
     \sigma_{23}&\sigma_{13}&\sigma_{34}^{\ast}&\sigma_{33}
\end{pmatrix}.
\end{eqnarray}
It is worthwhile to mention that the inputs of the covariance matrix are too long, and we avoid writing them out here. Note that
they can be obtained by performing the matrix product of the matrices 
Eqs. (\ref{rot}-\ref{haM}).
Furthermore, for isotropic  oscillators $R=1$, $ \Sigma(t,\Gamma) $  reduces to 
\begin{eqnarray}
\Sigma(t,\Gamma)=\begin{pmatrix}
     \sigma_{11}&\sigma_{12}&\sigma_{13}&\sigma_{23}^{\ast} \\
     \sigma_{12}^{\ast}&\sigma_{11}&\sigma_{23}&\sigma_{13}\\
     \sigma_{13}
     &\sigma_{23}^{\ast}&\sigma_{11}&\sigma_{12}\\
     \sigma_{23}&\sigma_{13}&\sigma_{12}^{\ast}&\sigma_{11}
\end{pmatrix}
\end{eqnarray}
where
the explicit expressions of the covariance inputs are 
	\begin{align}
	\sigma_{11}=& \frac{1}{2}{\sf{ch_1}}^2{\sf{sh_1}}^2\left(2-\exp\left[\Gamma t\left(\exp\left(\frac{2i\Omega_1}{\Gamma}\right)-1\right)\right]-\exp\left[\Gamma t\left(\exp\left(-\frac{2i\Omega_1}{\Gamma}\right)-1\right)\right]\right)
\\ \notag
&+\frac{1}{2}{\sf{ch_2}}^2{\sf{sh_2}}^2\left(2-\exp\left[\Gamma t\left(\exp\left(\frac{2i\Omega_2}{\Gamma}\right)-1\right)\right]-
\exp\left[\Gamma t\left(\exp\left(-\frac{2i\Omega_2}{\Gamma}\right)-1\right)\right]\right)+1\\
\sigma_{12}=& {\sf{ch_1}}^3{\sf{sh_1}}\left(-2+\exp\left[\Gamma t\left(\exp\left(\frac{2i\Omega_1}{\Gamma})\right)-1\right)\right]+
\exp\left[\Gamma t\left(\exp\left(-\frac{2i\Omega_1}{\Gamma}\right)-1\right)\right]\right)\\\notag
&+ {\sf{ch_2}}^3{\sf{sh_2}}\left(-2+\exp\left[\Gamma t\left(\exp\left(\frac{2i\Omega_2}{\Gamma}\right)-1\right)\right]+
\exp\left[\Gamma t\left(\exp\left(-\frac{2i\Omega_2}{\Gamma}\right)-1\right)\right]\right)\\ \notag 
&+ {\sf{ch_1}}{\sf{sh_1}}\left(1-\exp\left[\Gamma t\left(\exp\left(-\frac{2i\Omega_1}{\Gamma}\right)-1\right)\right]\right)
 + {\sf{ch_2}}{\sf{sh_2}}\left(1-\exp\left[\Gamma t\left(\exp\left(-\frac{2i\Omega_2}{\Gamma}\right)-1\right)\right]\right)\\
\sigma_{13}=& {\sf{ch_2}^{2}}{\sf{sh_2}^{2}}\left(2-\exp\left[\Gamma t\left(\exp\left(-\frac{2i\Omega_2}{\Gamma}\right)-1\right)\right]-
\exp\left[\Gamma t\left(\exp\left(\frac{2i\Omega_2}{\Gamma}\right)-1\right)\right]\right)\\ \notag
&+\sf{{ch_1}^{2}}{\sf{sh_1}^{2}}\left(-2+\exp\left[\Gamma t\left(\exp\left(-\frac{2i\Omega_1}{\Gamma}\right)-1\right)\right]+
\exp\left[\Gamma t\left(\exp\left(\frac{2i\Omega_1}{\Gamma}\right)-1\right)\right]\right)\\ 
\sigma_{23}=& {\sf{ch_1}}^3{\sf{sh_1}}\left(2-\exp\left[\Gamma t\left(\exp\left(\frac{2i\Omega_1}{\Gamma}\right)-1\right)\right]-\exp\left[\Gamma t\left(\exp\left(-\frac{2i\Omega_1}{\Gamma}\right)-1\right)\right]\right)\\\notag
&+ {\sf{ch_2}}^3{\sf{sh_2}}\left(-2+\exp\left[\Gamma t\left(\exp\left(\frac{2i\Omega_2}{\Gamma}\right)-1\right)\right]+\exp\left[\Gamma t\left(\exp\left(-\frac{2i\Omega_2}{\Gamma}\right)-1\right)\right]\right)\\ \notag 
&- {\sf{ch_1}}{\sf{sh_1}}\left(1-\exp\left[\Gamma t\left(\exp\left(\frac{2i\Omega_1}{\Gamma}\right)-1\right)\right]\right)
 + {\sf{ch_2}}{\sf{sh_2}}\left(1-\exp\left[\Gamma t\left(\exp\left(\frac{2i\Omega_2}{\Gamma}\right)-1\right)\right]\right)\notag 
	\end{align}
 where $\sf{ch_j}=\cosh(s_j)$ and $\sf{sh_j}=\sinh(s_j)$ have been defined. The virtual excitations in  both modes are 
\begin{eqnarray}
\langle N_1\rangle=\langle a^{+}a\rangle= \tfrac{1}{2}\left(\sigma_{11}-1\right), \qquad  \langle N_2\rangle=\langle b^{+}b\rangle= \tfrac{1}{2}\left(\sigma_{22}-1\right).
\end{eqnarray}
It is  straightforward to show that the virtual excitations  in both oscillators are equal and reduce to
\begin{align}
\langle a^{+}a\rangle(t)=\langle b^{+}b\rangle(t)=&\frac{1}{2}{\sf{ch_1}}^2{\sf{sh_1}}^2\left(2-\exp\left[\Gamma t\left(\exp\left(\frac{2i\Omega_1}{\Gamma}\right)-1\right)\right]-\exp\left[\Gamma t\left(\exp\left(-\frac{2i\Omega_1}{\Gamma}\right)-1\right)\right]\right)
\\ \notag
&+\frac{1}{2}{\sf{ch_2}}^2{\sf{sh_2}}^2\left(2-\exp\left[\Gamma t\left(\exp\left(\frac{2i\Omega_2}{\Gamma}\right)-1\right)\right]-\exp\left[\Gamma t\left(\exp\left(-\frac{2i\Omega_2}{\Gamma}\right)-1\right)\right]\right).
\end{align}
Both the single reduced modes and the correlation sub-matrix will be derived by using the trace-out prescription and, as a consequence, we obtain
\begin{eqnarray}
\Sigma^{(a)}=\Sigma^{(b)}=\left(\begin{array}{cc}
     \sigma_{11}&  \sigma_{12}\\
    \sigma_{12}^{\ast} & \sigma_{11}
\end{array}\right), \qquad \Sigma^{(a  b)}=\left(\begin{array}{cc}
     \sigma_{13}&  \sigma_{23}^\ast\\
    \sigma_{23} & \sigma_{13}
\end{array}\right).
\end{eqnarray}

To evaluate the entanglement, we compute the partial transposition $\tilde{\Sigma}$ and obtain the four  eigenvalues 
\begin{eqnarray}
\tilde{\nu}_{m,M}^{2}= \frac{1}{2}\left(\delta\mp\sqrt{\delta^{2}-4\det \Sigma}\right)
\end{eqnarray}
where the seralian $\delta$ is defined by
\begin{eqnarray}
\delta=\det\Sigma^{(a)}+\det\Sigma^{(b)}-2\det\Sigma^{(ab)}.
\end{eqnarray}
Consequently, the logarithmic negativity becomes
\begin{eqnarray}
E_{N}=\max\left(0,-\ln(\tilde{\nu}_{m})\right).
\end{eqnarray}
Distinct from entanglement, quantum steering is a magic quantum correlation that is generally asymmetric. 
 Thus,  one needs  to define the steering in both ways \cite{steering}
\begin{eqnarray}
S^{A\rightarrow B}=\max\left(0,\log_{2}\sqrt{\frac{\det \Sigma^{(a)}}{4\det \Sigma}}\right), \qquad S^{B\rightarrow A}=\max\left(0,\log_{2}\sqrt{\frac{\det \Sigma^{(b)}}{4\det \Sigma}}\right)
\end{eqnarray}
and  the steering asymmetry is
\begin{eqnarray}
\Delta S=|S^{A\rightarrow B}-S^{B\rightarrow A}|. \label{asymmetry}
\end{eqnarray}
We mention that the steering is symmetric in the isotropic case since
$S^{A\rightarrow B}=S^{B\rightarrow A}$.

	\section{\label{sec4}Numerical results}
	\subsection{Avoiding intrinsic decoherence}
		In Fig. \ref{fig4}, we show in a comparative study the relationship between Milburn and von Neumann dynamics. For this reason, we plot the evolution of the virtual excitations for different values of ultra-strongly coupling $J$.  As expected, the dynamics of  the virtual excitations  under von Neumann dynamics exhibit an unquenched oscillatory behavior. We set the decoherence rate $\Gamma=100$, the excitations undergo a quenched oscillatory behavior that becomes very slow with the increase of ultra-strongly coupling. Subsequently, near the hermitianity point where the ultra-strong coupling reaches its maximum, the dynamics overlap each other. Therefore, we conclude that the intrinsic decoherence of a quantum system can be removed by ultra-strongly coupling its internal degrees of freedom.
		
	\begin{figure}[H]
	\centering
	\includegraphics[width=5cm, height=3.8cm]{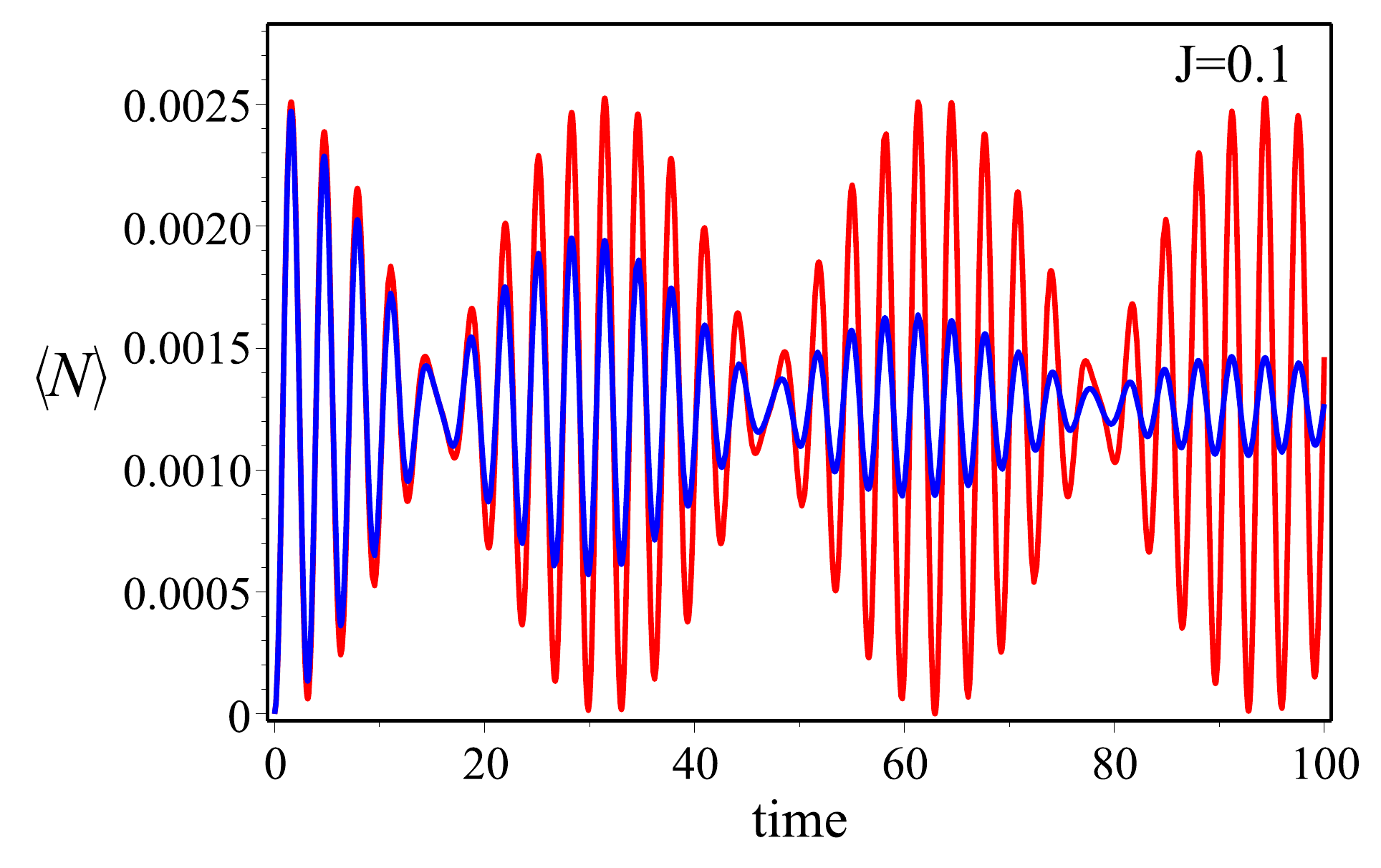}
	\includegraphics[width=5cm, height=3.8cm]{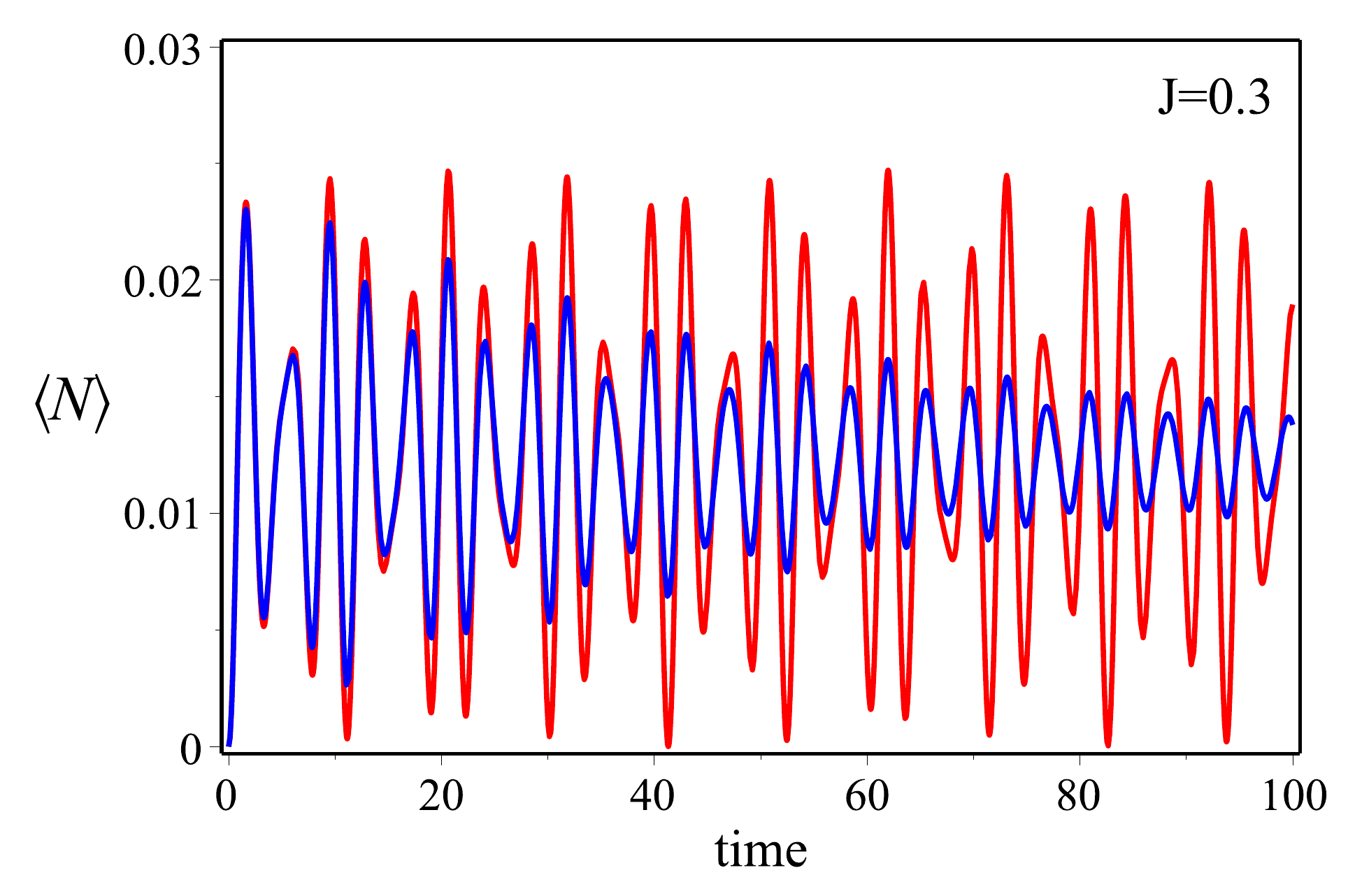}
	\includegraphics[width=5cm, height=3.8cm]{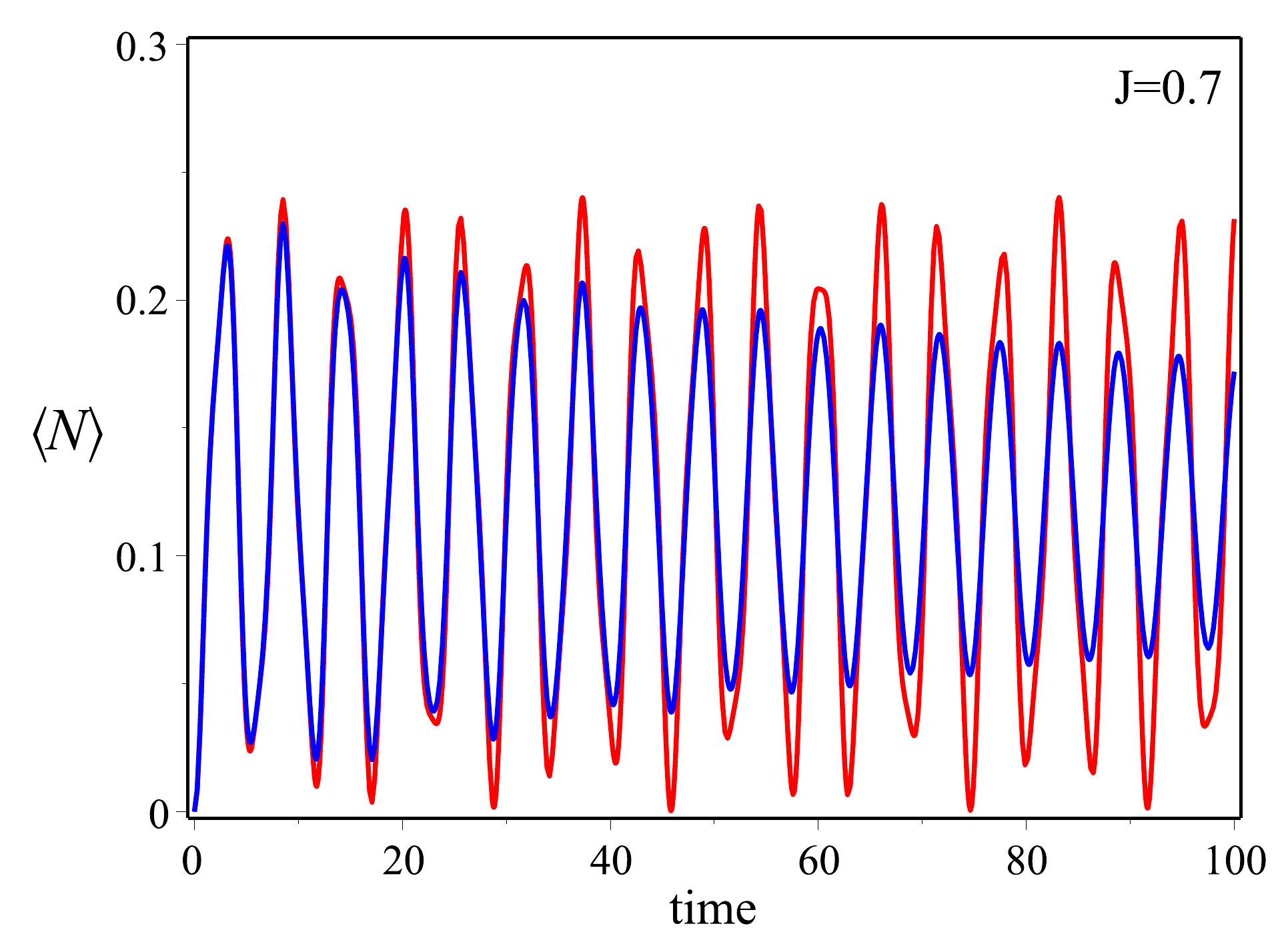}
	\includegraphics[width=5cm, height=3.8cm]{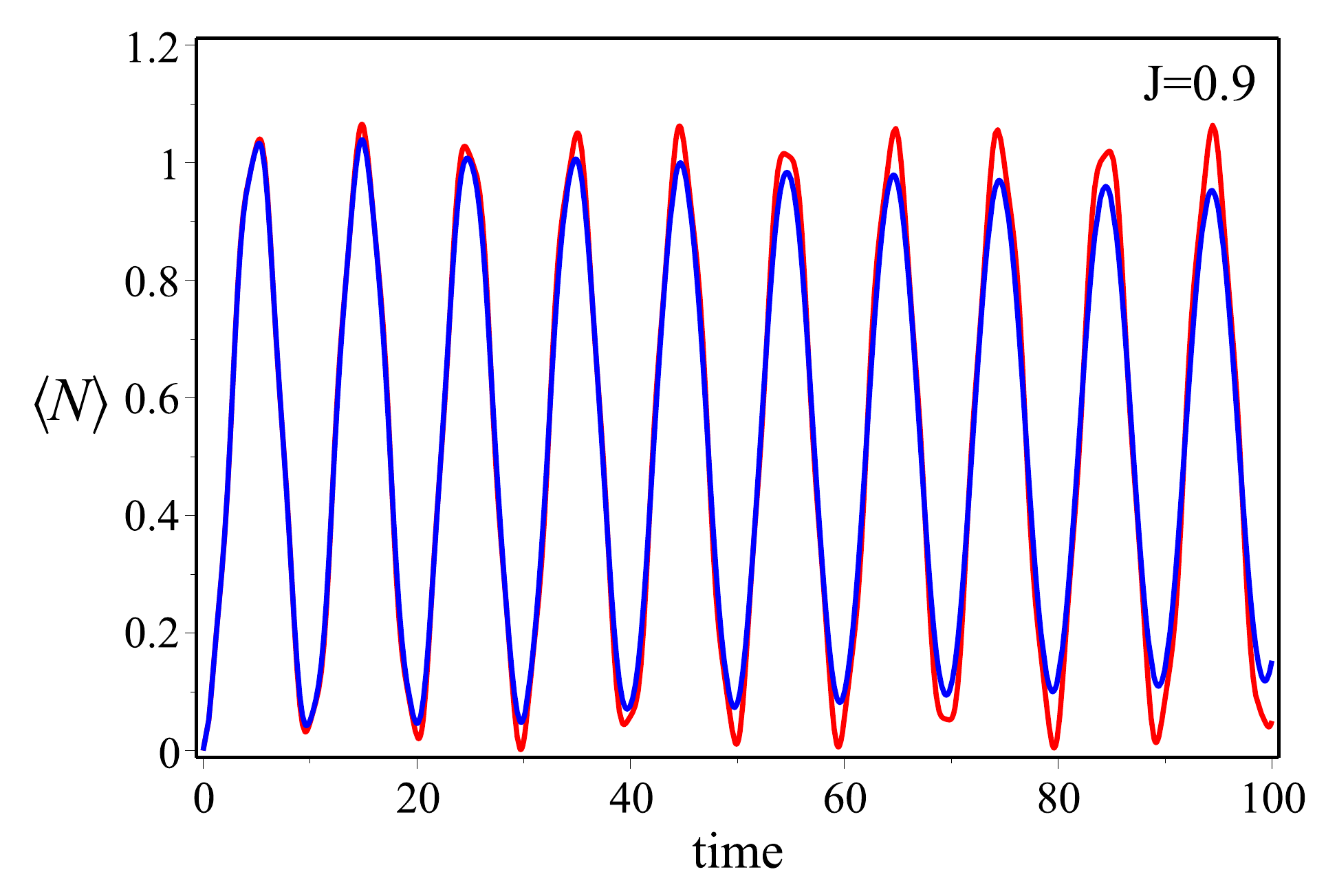}
	\includegraphics[width=5cm, height=3.8cm]{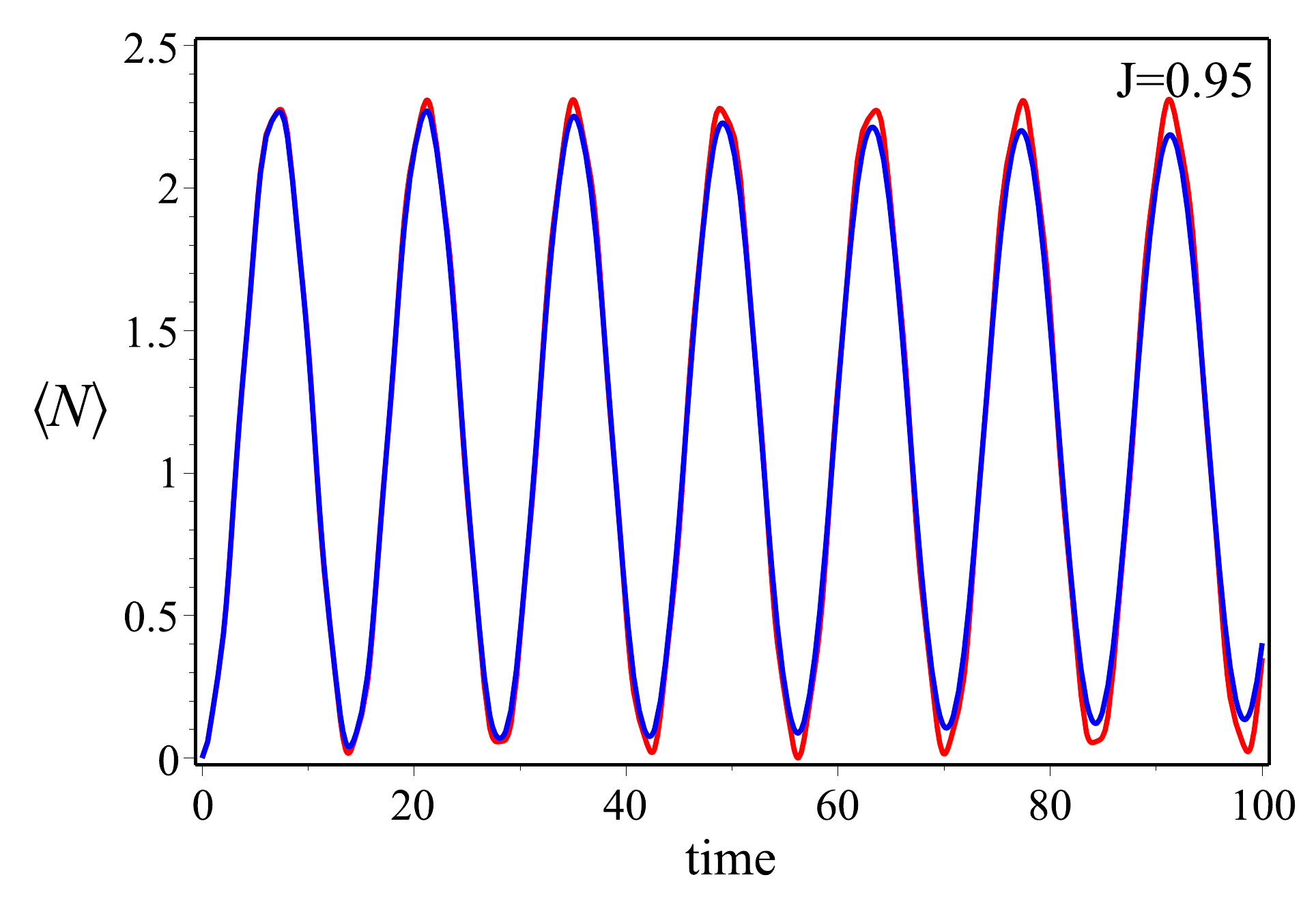}
	\includegraphics[width=5cm, height=3.5cm]{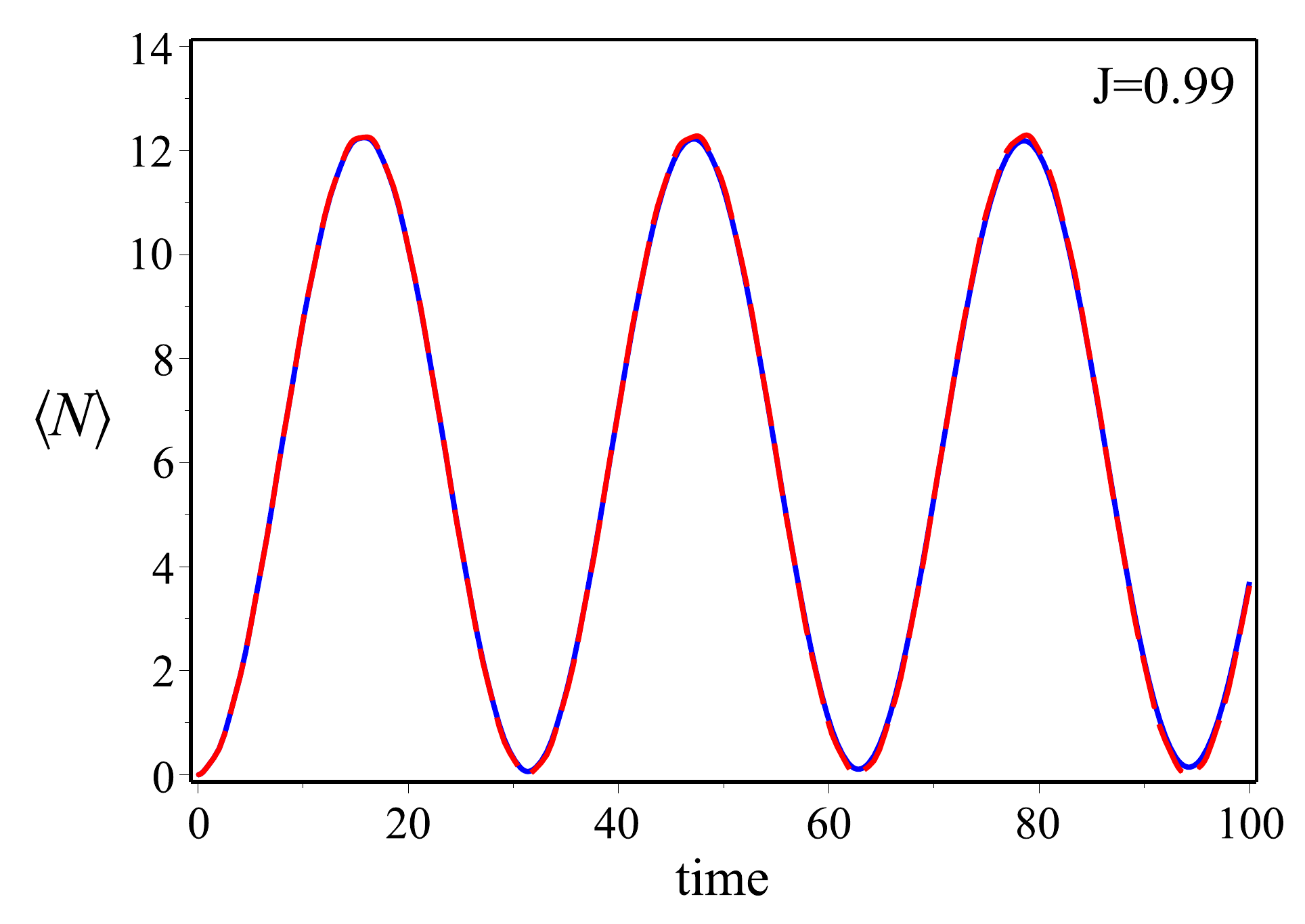}
	\caption{(color online) Milburn and von Neumann dynamics under an ultra-strongly coupling effect with the blue line represents von Neumann dynamics and the red line, Milburn dynamics.  We choose $\Gamma=100$ and $\omega_1=\omega_2=1$. }\label{fig4}
\end{figure}

			\subsection{Role of anisotropy}
	
		In 	Fig. \ref{fig2c}, we  investigate the behaviour of quantum entanglement $E_N$, steering ($S^{A\rightarrow B}$ and $S^{B\rightarrow A}$) and  virtual excitations ($\langle N\rangle_1$ and $\langle N\rangle_2$) versus scaled time for
		$\Gamma=100$, $J=0.2$. Before analyzing our results, we recall that for weak coupling, the ground state is empty and since the coupling takes important values, namely, the ultra-strong coupling regime, the ground state becomes populated. According to our previous results  \cite{quantum reports}, these excitations are linked to the classical instability of classical oscillators and maintain entanglement between oscillators. Now,  in the first line of figures we address the dynamics in the resonant case $\omega_1=\omega_2=1$. Excitations and one-way steering will be the same, as expected. Subsequently, we observe that dynamics  rapidly revives excitations that will maintain entanglement and steering between oscillators. In addition, we mention that due to intrinsic decoherence, the three quantities undergo an exponential decay to a steady value except for steering which suddenly dies out. This shows that entanglement and excitations are robust against decoherence. This profile is  mathematically due to the intrinsic decoherence factor  $\exp[-\Gamma t]$. Subsequently, we slightly move away from resonance by taking $\omega_2=0.95$ and maintaining  all other parameters. At first glance, we amazingly notice the  redistribution of excitations as well as their distinguishability    $D=\langle N_1 \rangle-\langle N_2 \rangle \neq 0$. As a result, the asymmetry  in quantum  steering $\Delta S\neq 0$  and a soft decrease in entanglement compared to the former case. Thirdly, we vary $\omega_2$ to $0.7$ and $0.3$, the distinguishability  $D$ increases, and as a consequence, the profile of entanglement and steering shows enhancement immediately after their generation, but because of decoherence, they fastly die out. 
		For completeness, reduce $\omega_2$ to $0.21$ and observe that excitations result in a quantum synchronisation regime \cite{synchro}.
%
As a result, the quantum correlations are significantly enhanced and revived for a long time.
Finally, it is worthwhile to mention the hierarchy of quantum correlations \cite{hierarchy} and excitations. From the above results, we have 
	  	$ \langle N_1 \rangle=\langle N_2 \rangle=0$ showing  $E_N=0 $ and therefore $ S^{A\rightarrow B}=S^{B\rightarrow A}=0 $.
 This hierarchy relation means, for instance, that the establishment of correlations necessarily entails the appearance of excitations. This will allow a new route to construct new quantifiers based on  excitations \cite{maintaining,quantum reports}.
 Another issue that deserves attention is the dependence of the one-way steerability on the angular velocity $ \omega_{1,2} $. More clearly, the steerabilty of  a rapid oscillator from a slow one always exceeds the inverted steerabilty. This may be due to the obvious distinguishability of excitations. 
	To conclude, moving away from resonance (increasing  the anisotropy $R=\sqrt{{\omega_1}/{\omega_2}}$) has the following implications: (i) A generation
		   of excitations followed by a  redistribution  phenomenon. A similar phenomenon was observed in a system made of two oscillators interacting with a qubit \cite{qubit and synchr}.
		   (ii) The  redistribution of excitations implies the redistribution of quantum steering. For instance, $S^{A\rightarrow B}<S^{B\rightarrow A}$ if only and if $\langle N_1 \rangle <\langle N_2 \rangle$, this is consistent with recent results of magnon-photon steering in ferromagnetic and anti-ferromagnetic sub-lattices \cite{magnon}.
		   (iii) The activation of virtual excitation generation, which results in the preservation and enhancement of quantum correlations between oscillators. (iv) The system enters its quantum synchronous regime, at which point excitations and correlations become significant. This is similar to what has been discovered in other works, such as the qubit-qubit system
		   \cite{qubit-qubit,Qubit-Qubit}   and oscillators \cite{oscil1,oscil2}.
		   	\begin{figure}[H]
		\centering
		\includegraphics[width=5cm, height=4cm]{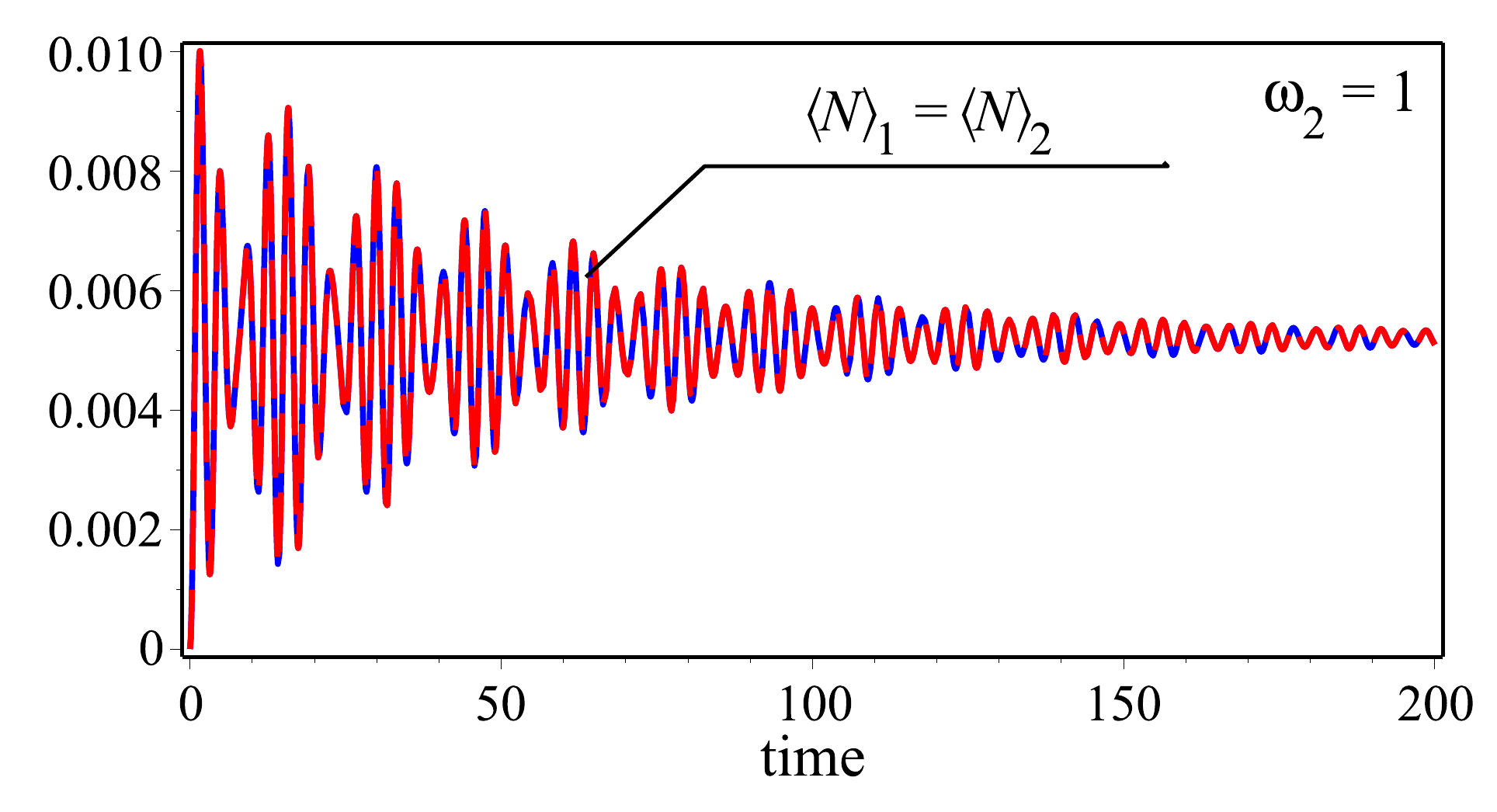}
		\includegraphics[width=5cm, height=3.8cm]{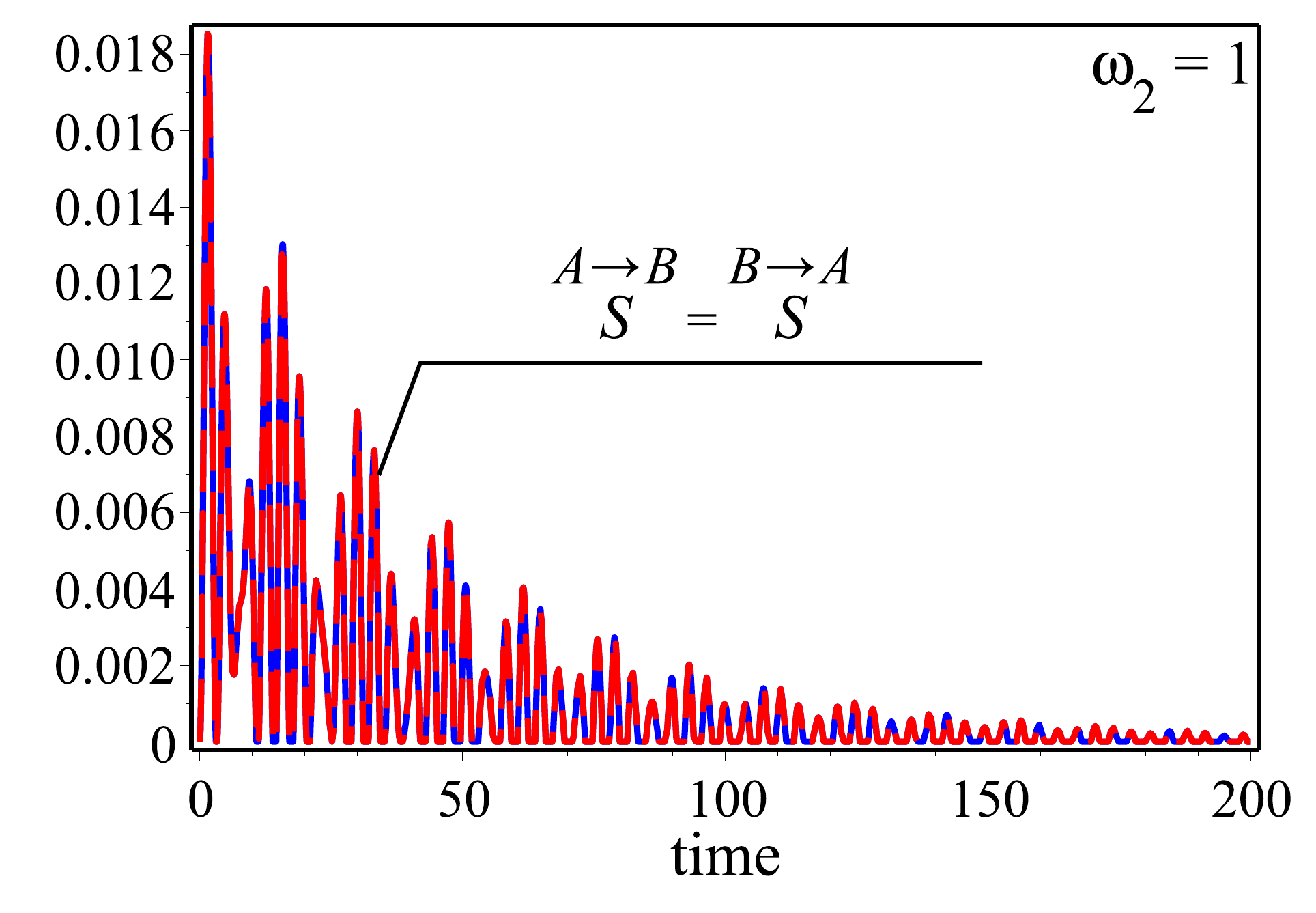}
		\includegraphics[width=5cm, height=3.8cm]{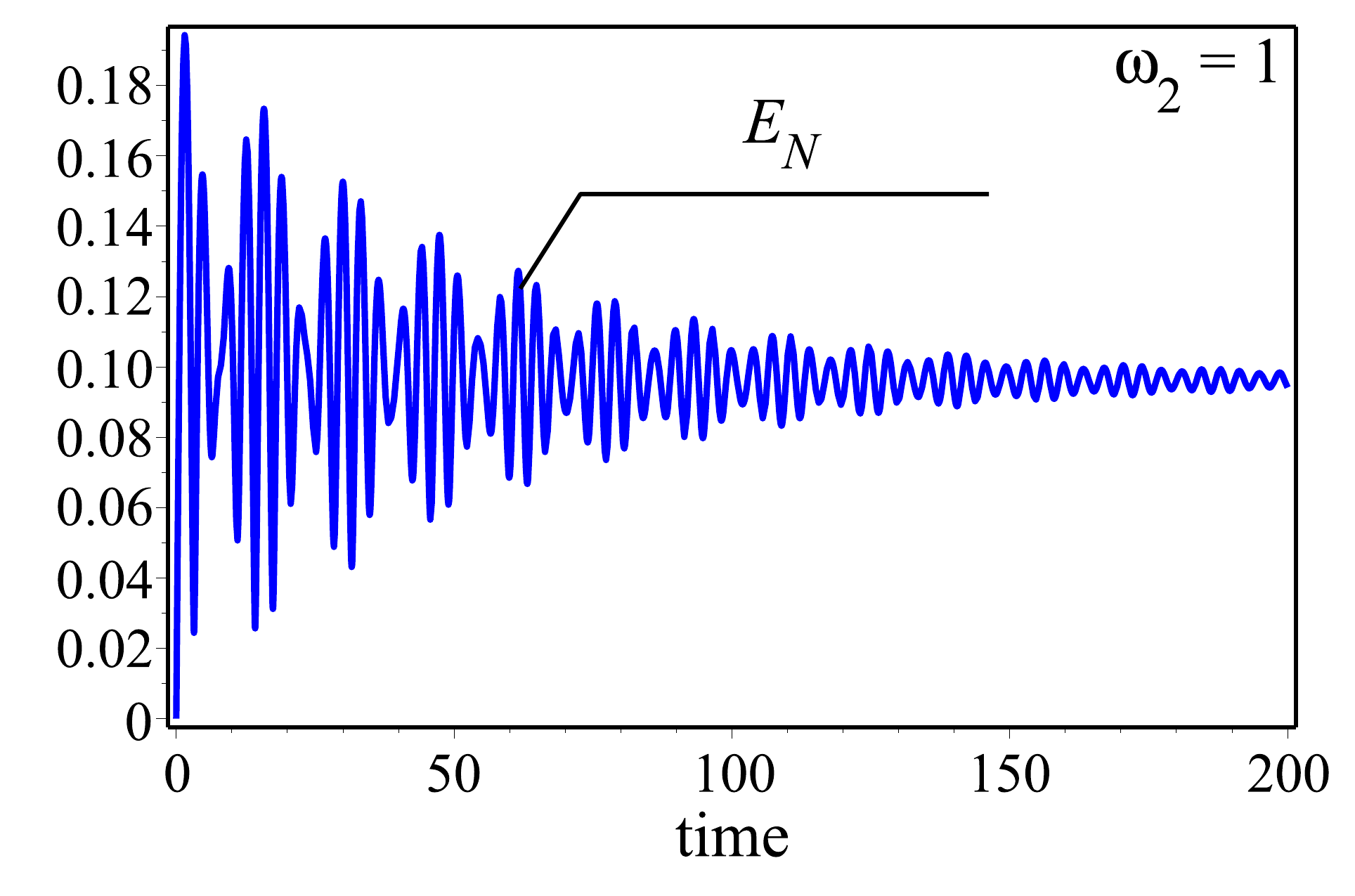}
		\includegraphics[width=5cm, height=3.8cm]{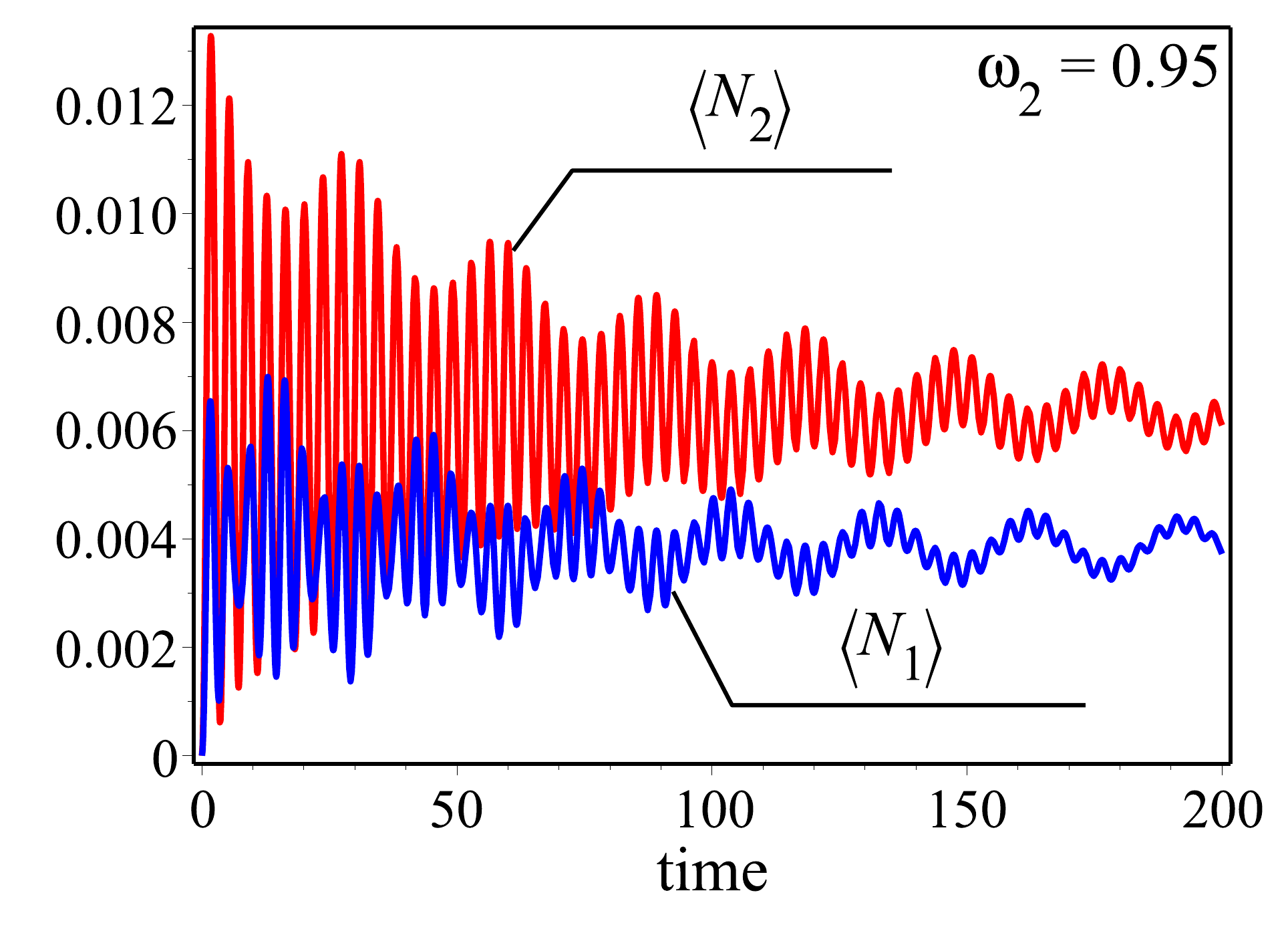}
		\includegraphics[width=5cm, height=3.8cm]{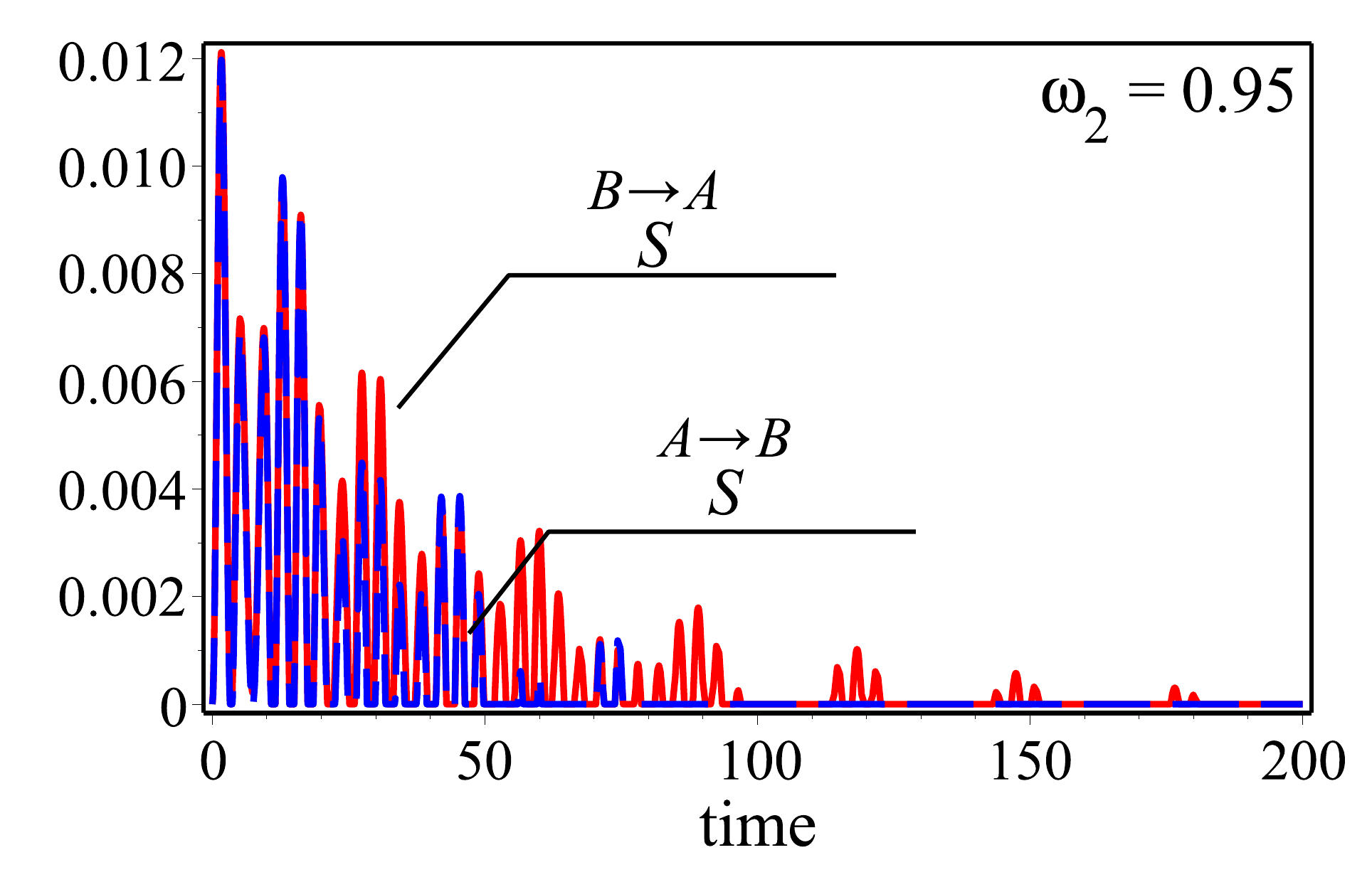}
		\includegraphics[width=5cm, height=3.8cm]{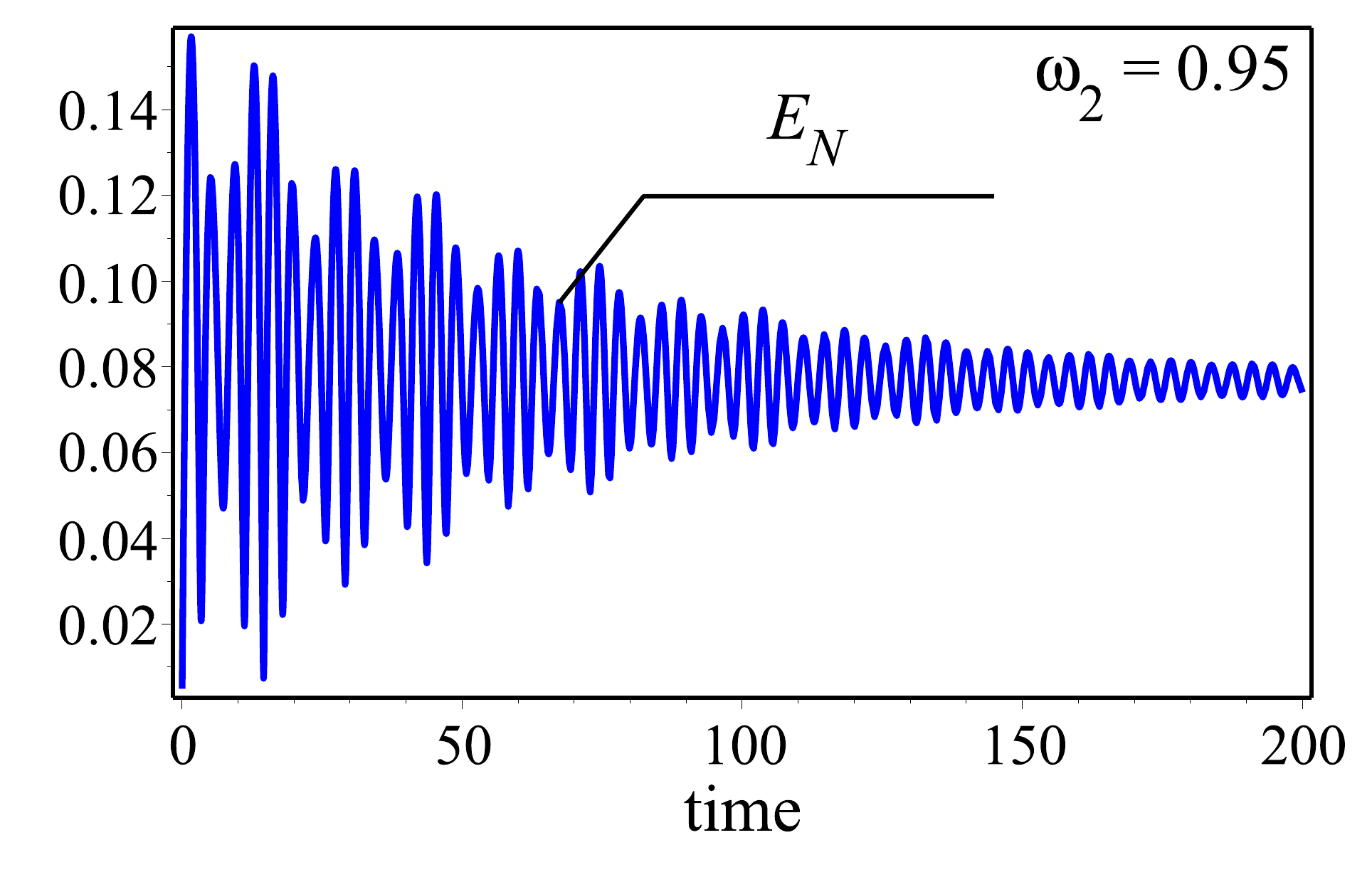}	
		\includegraphics[width=5cm, height=3.8cm]{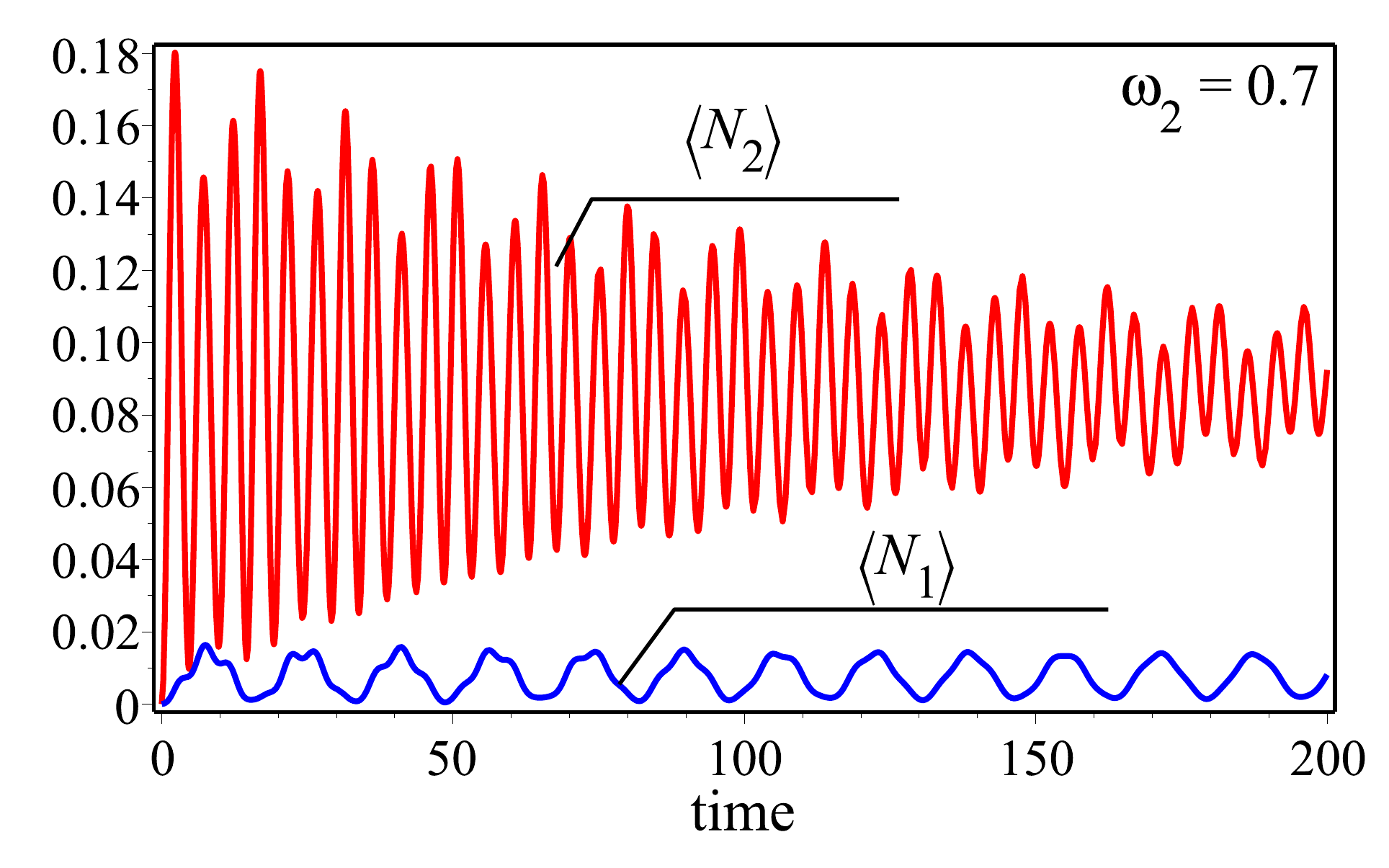}
		\includegraphics[width=5cm, height=3.8cm]{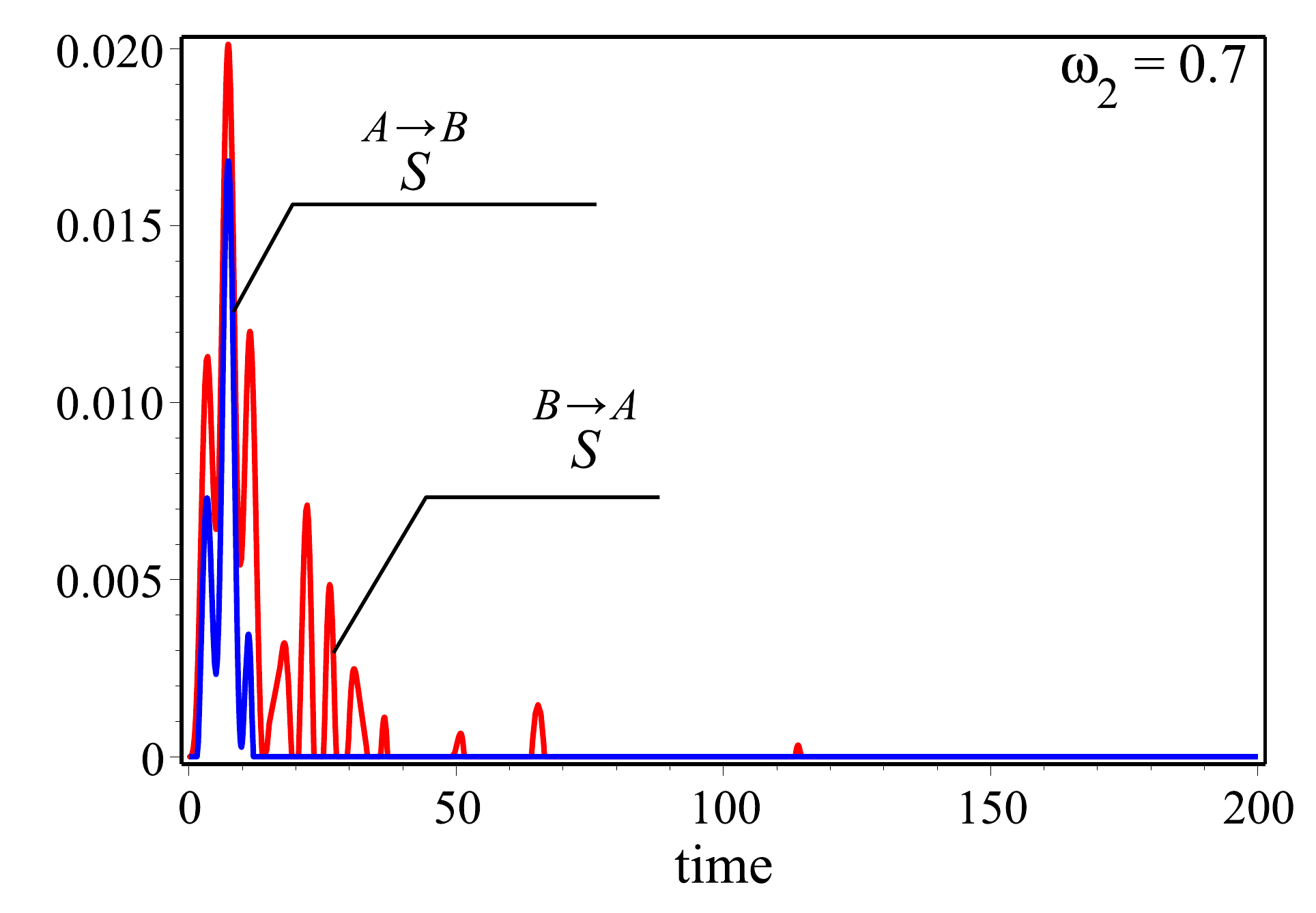}
		\includegraphics[width=5cm, height=3.8cm]{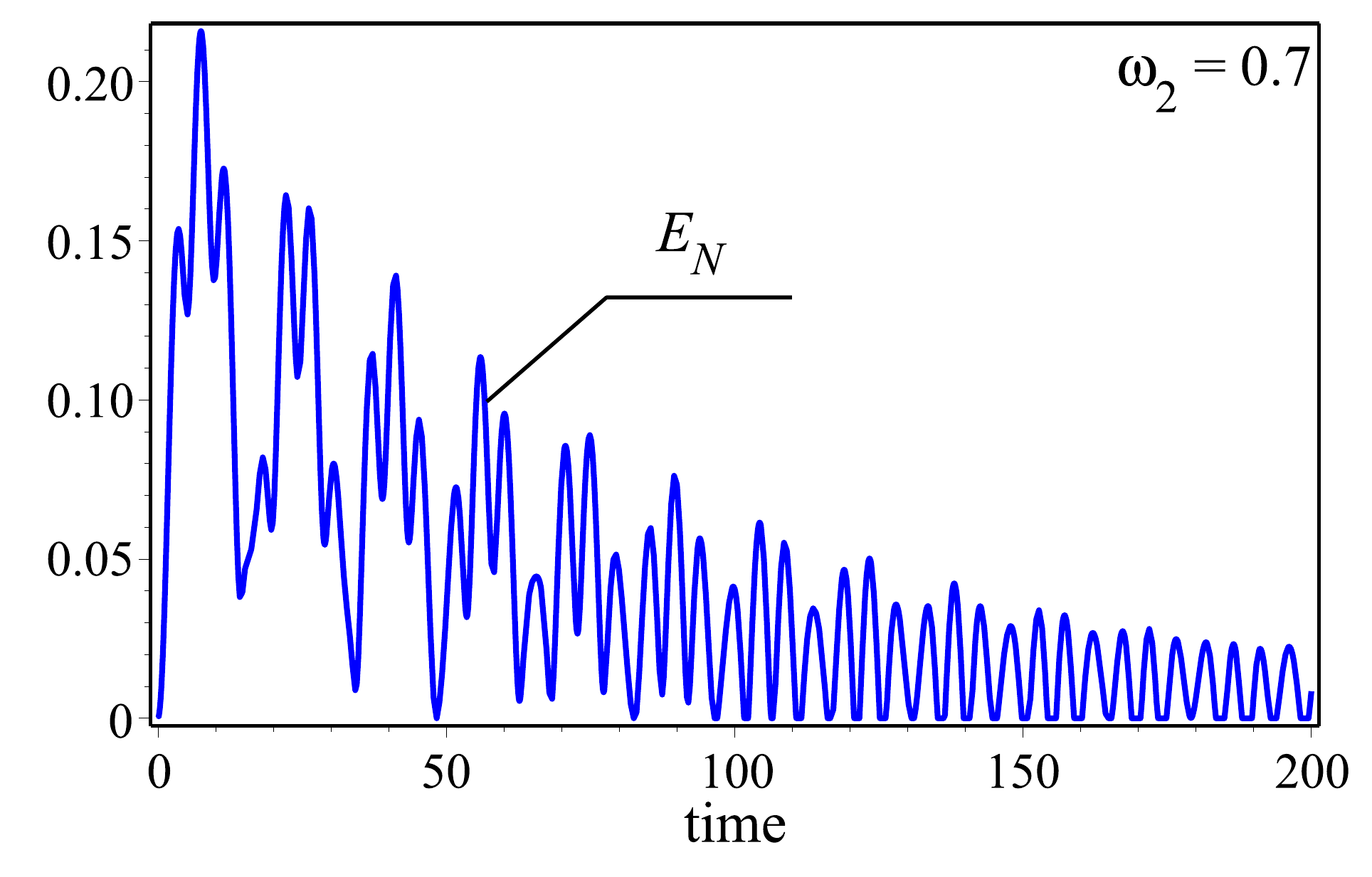}
		\includegraphics[width=5cm, height=3.8cm]{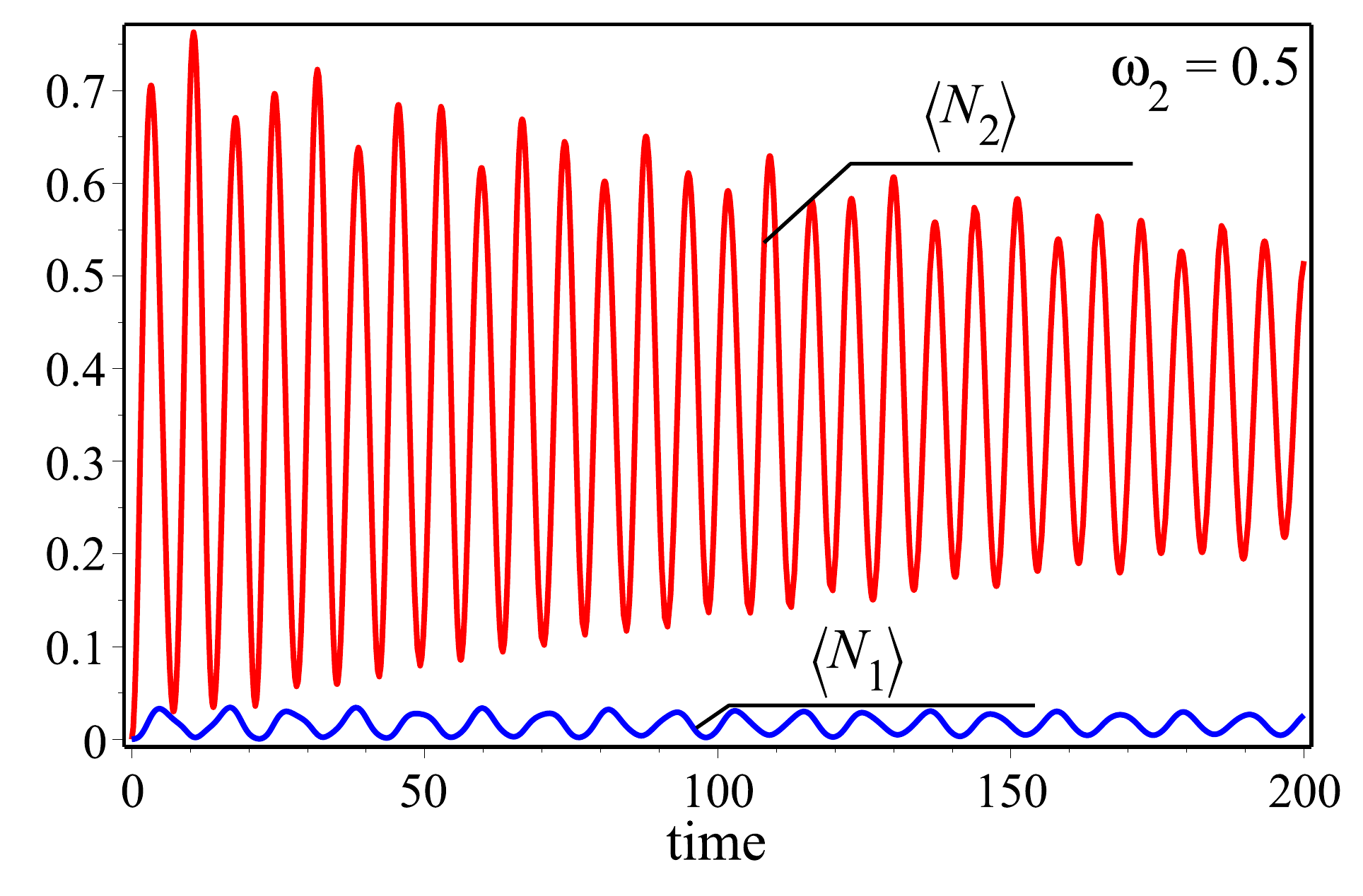}
		\includegraphics[width=5cm, height=3.8cm]{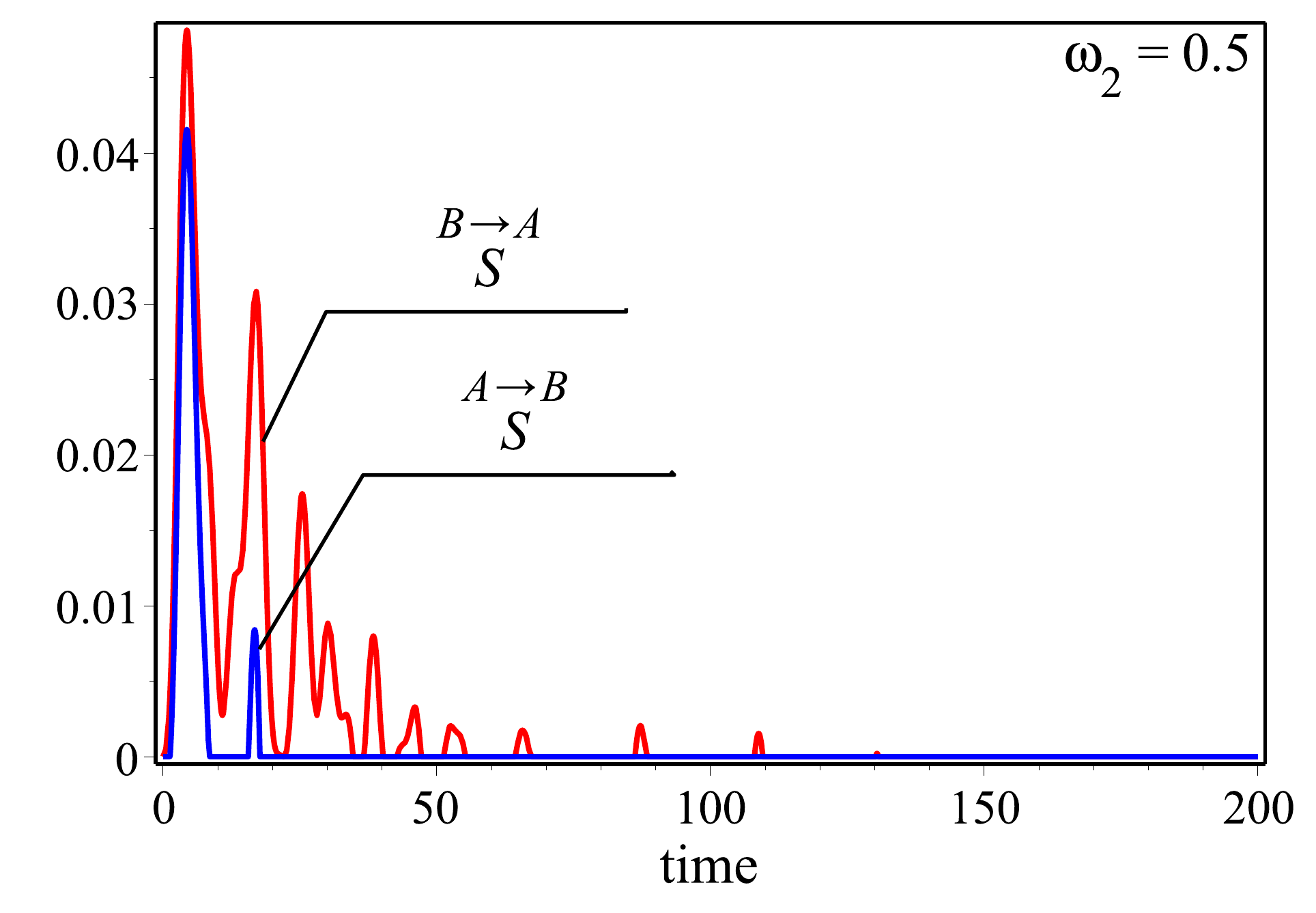}
		\includegraphics[width=5cm, height=3.8cm]{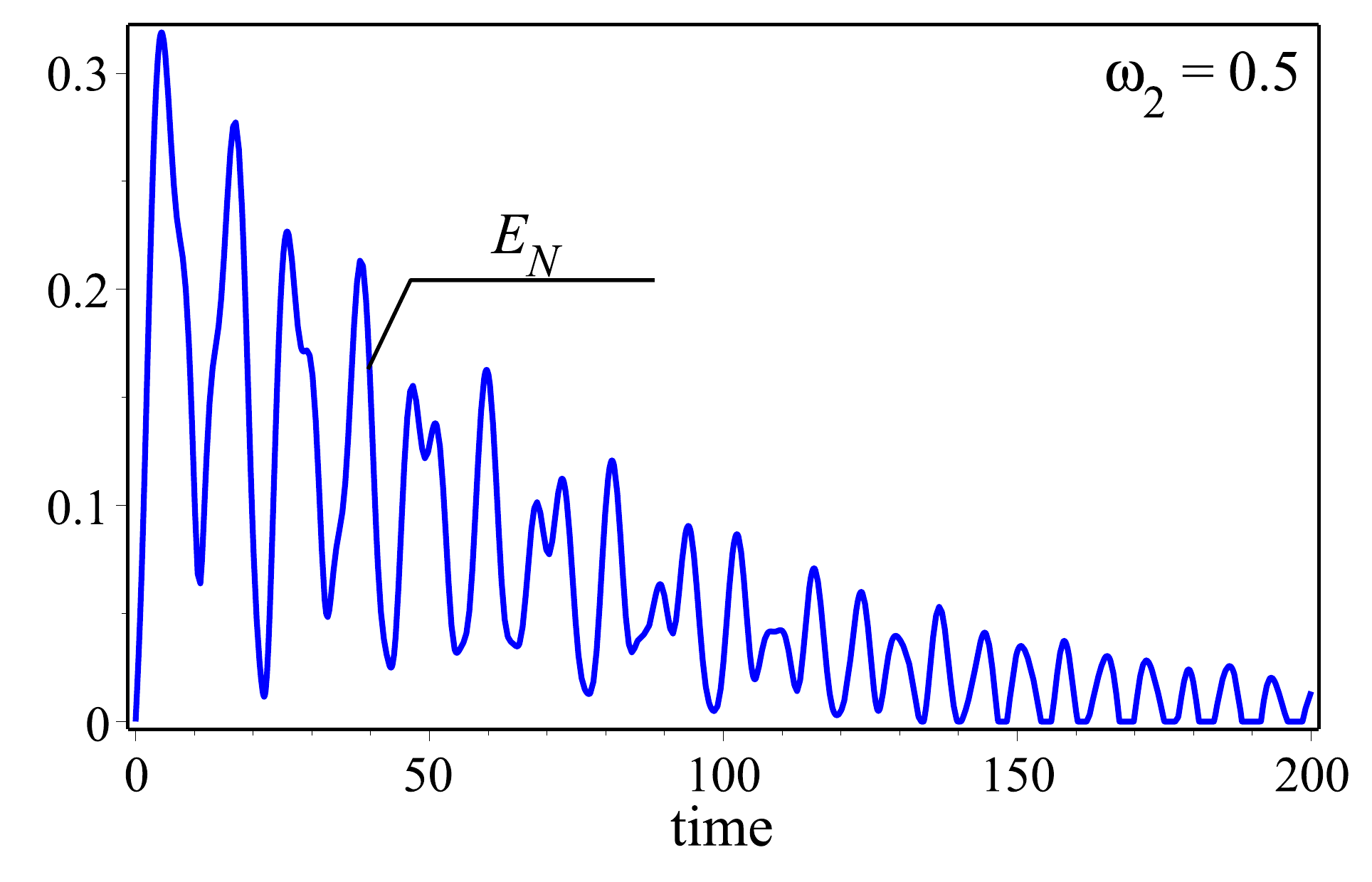}	\includegraphics[width=5cm, height=3.8cm]{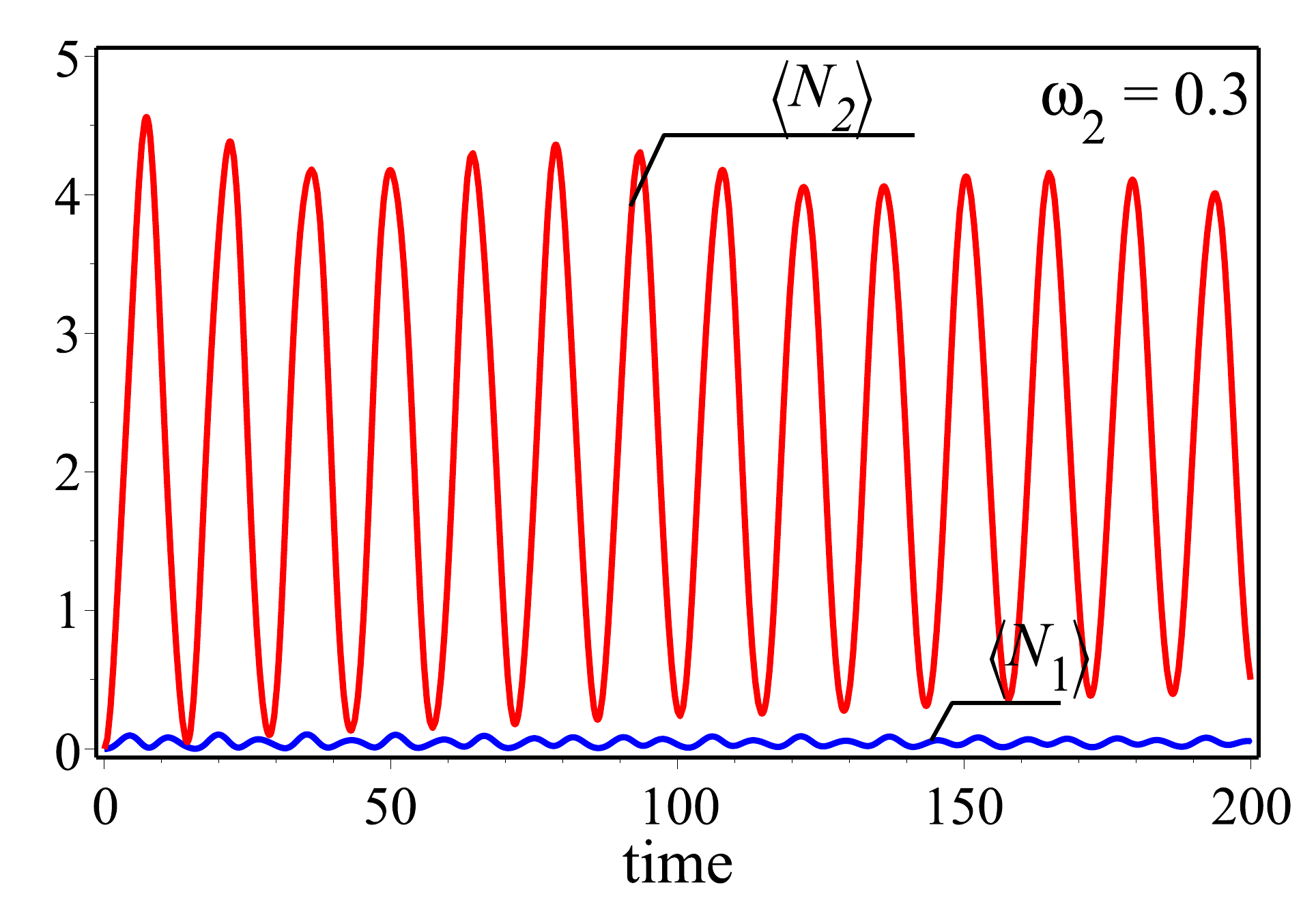}
		\includegraphics[width=5cm, height=3.8cm]{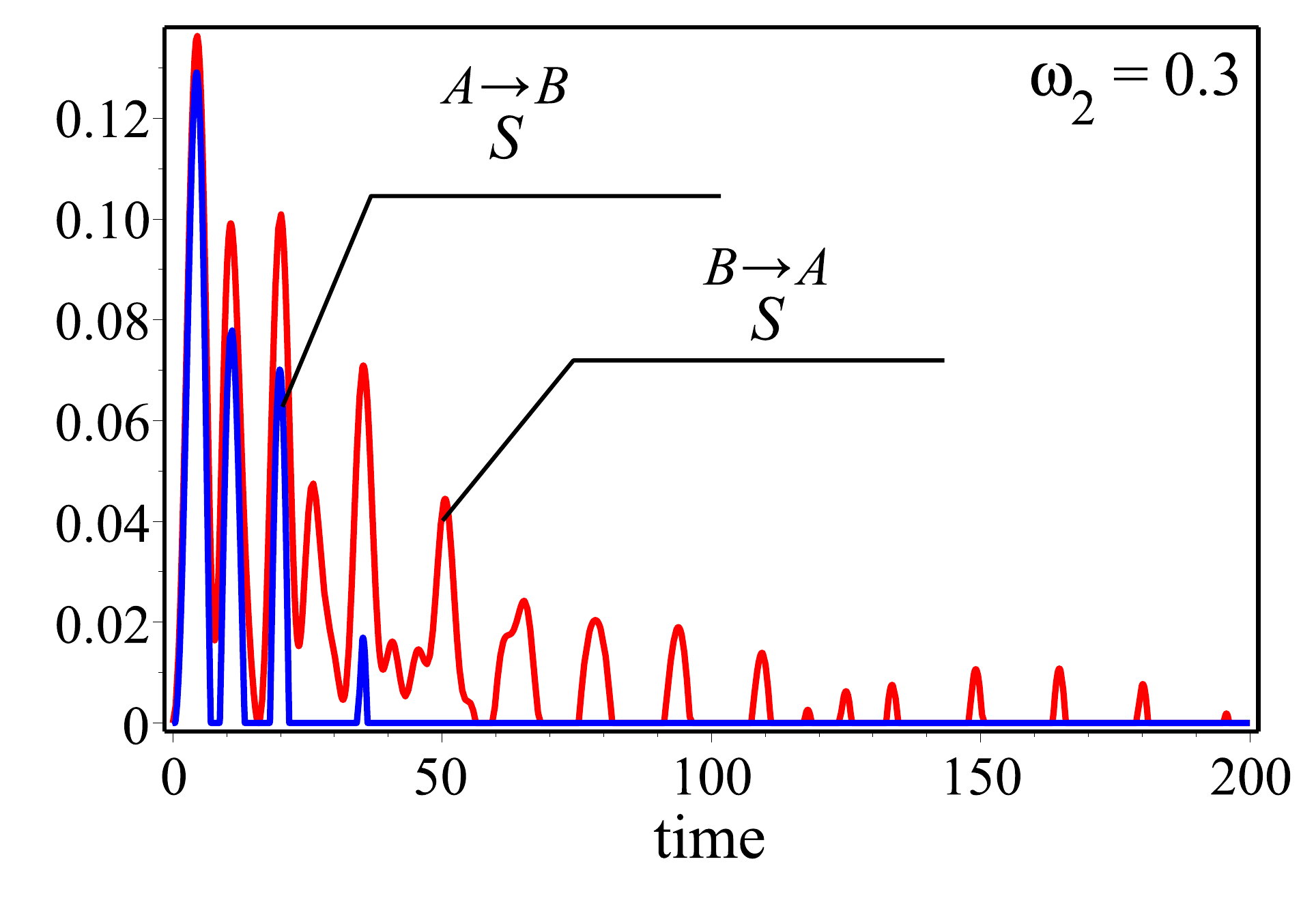}
		\includegraphics[width=5cm, height=3.8cm]{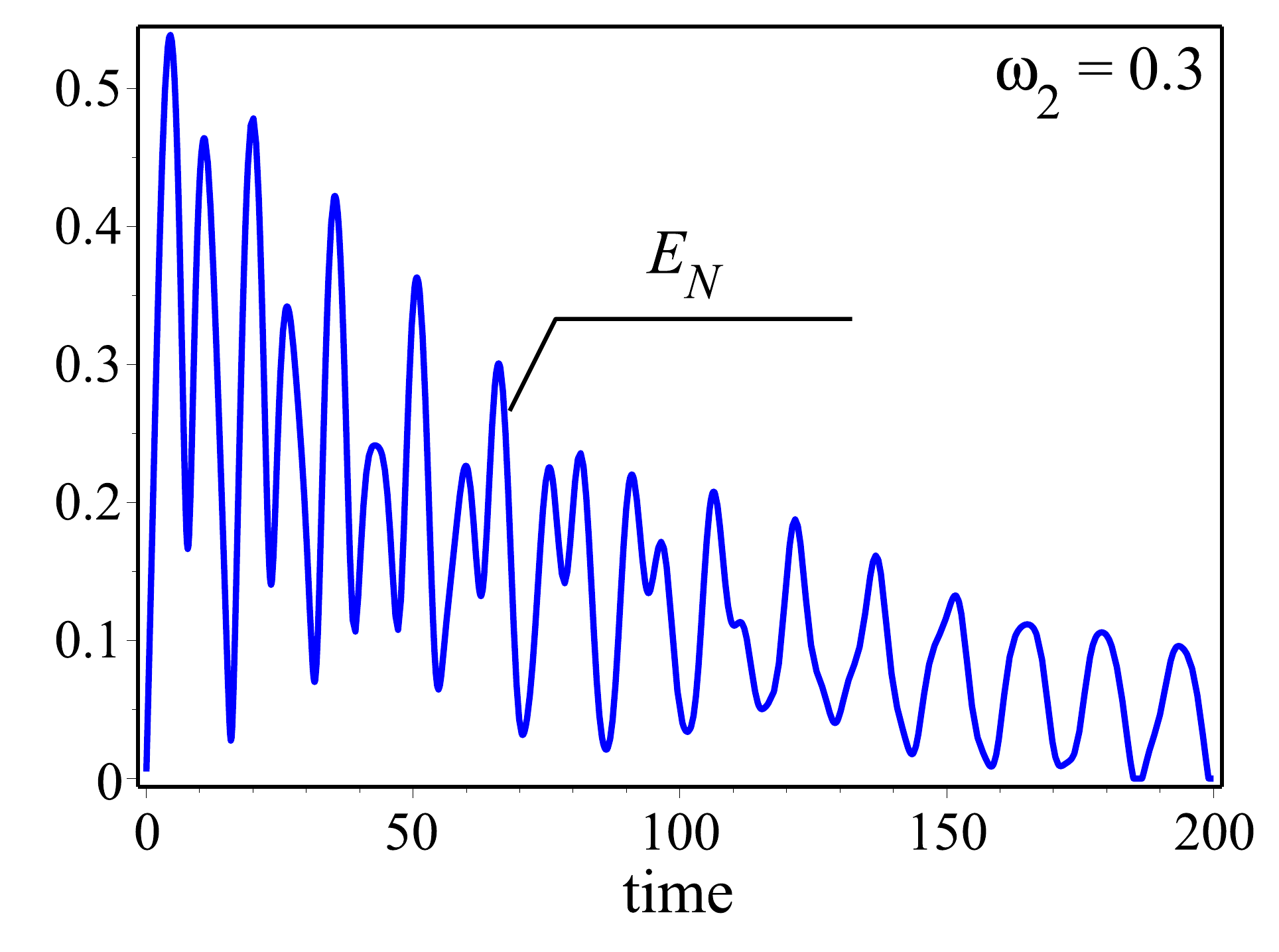}
		\includegraphics[width=5cm, height=3.8cm]{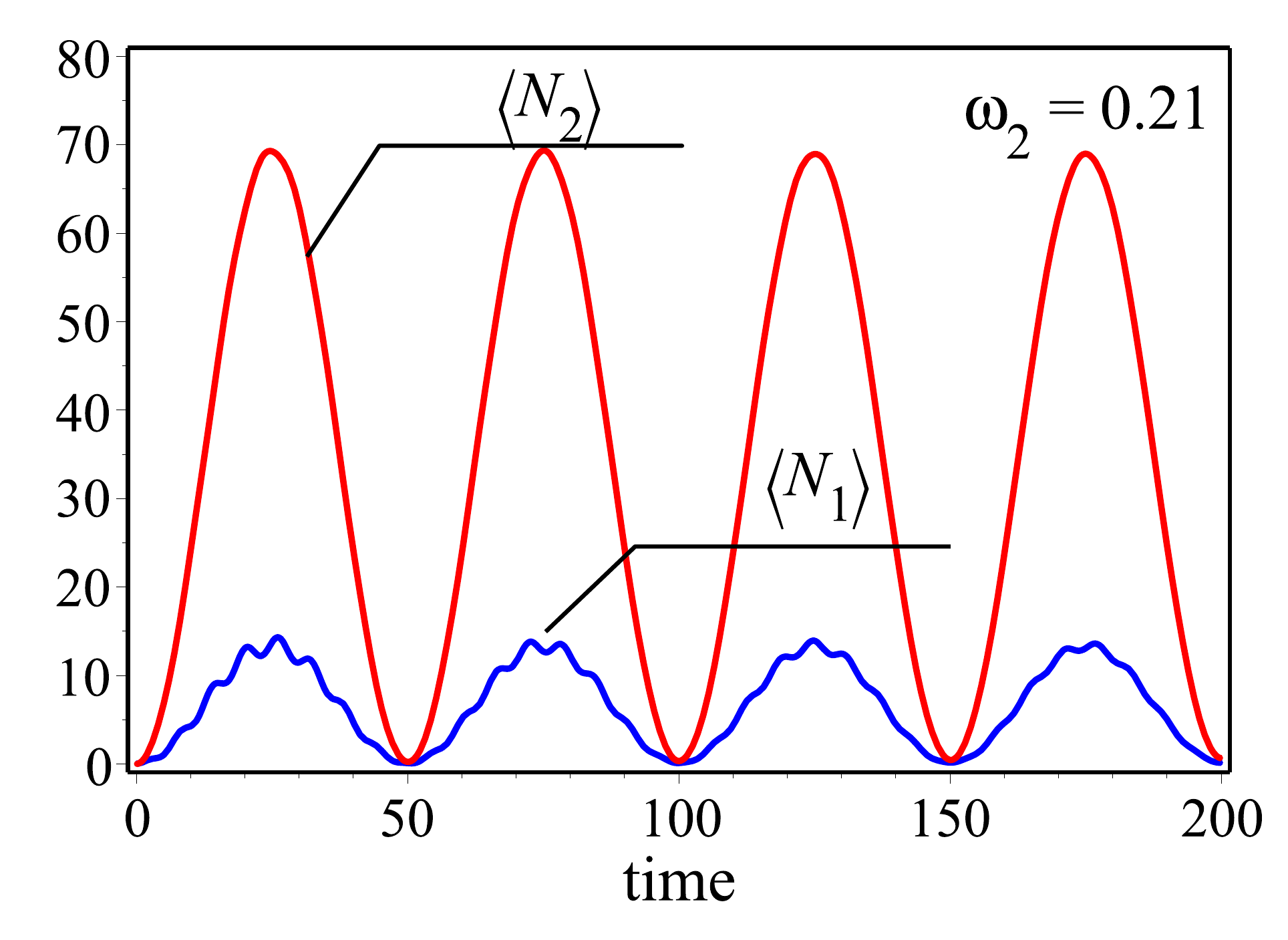}
		\includegraphics[width=5cm, height=3.8cm]{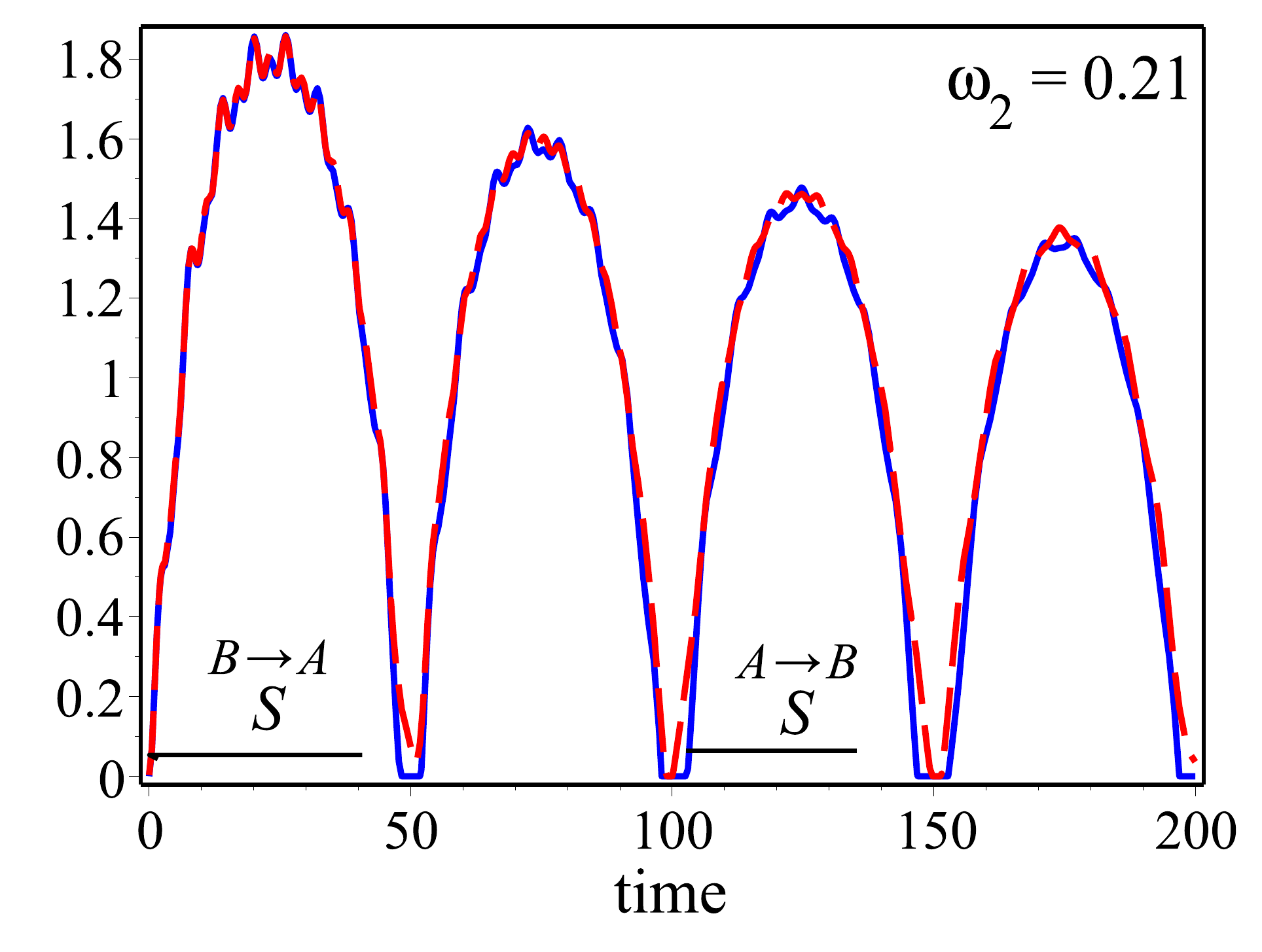}
		\includegraphics[width=5cm, height=3.8cm]{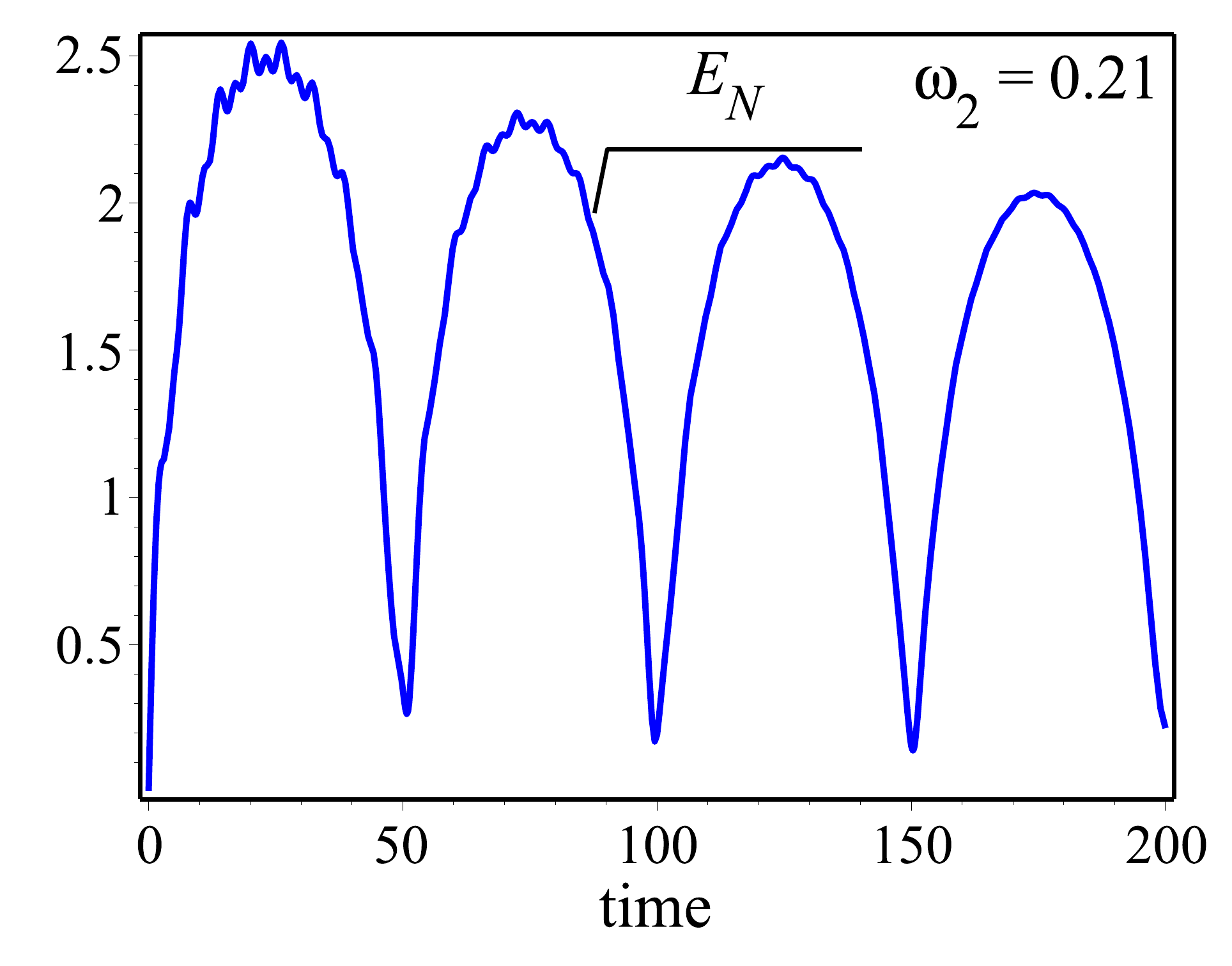}	
		\caption{(color online) In the first column, we plot the dynamics of excitations in both oscillators. In the second column, we plot the dynamics of quantum steering for both cases. In the third one, we present the dynamics of entanglement. We choose  $\Gamma=100$, $\omega_1=1$ and  $J=0.2$. }\label{fig2c}
	\end{figure}
	
		   	\subsection{Effect of ultra-strong coupling}
	 In	Fig.  \ref{fig3c},  we  investigate the dynamics of the three quantities and their interconnection by varying  the ultra-strong coupling $J$ for  $\Gamma=100$,  $\omega_1=1$, $\omega_2=0.5$. We start with $J=0.1$, the distinguishability is evident and $\langle N_1\rangle < \langle N_2\rangle$ for all time. As a result, the one-way steering $S^{A\rightarrow B}$ is more fragile than the second steering $S^{B\rightarrow A}$. By gradually  increasing the ultra-strong coupling, we observe that the excitations are significantly generated, as well as a revival of one-way steering $S^{A\rightarrow B}$. 
	 The two oscillators will become synchronous and the quantum correlations will be greatly enhanced with a large coupling
	 $J\rightarrow J_{max}=\sqrt{\omega_1\omega_2}$.  To summarize, a large coupling constant $J$ combined with a large anisotropy constant $R$ can result in a significant number of quantum correlations even in the presence of decoherence and over long simulation times. 
	 	\begin{figure}[H]
		\centering	\includegraphics[width=5cm, height=4cm]{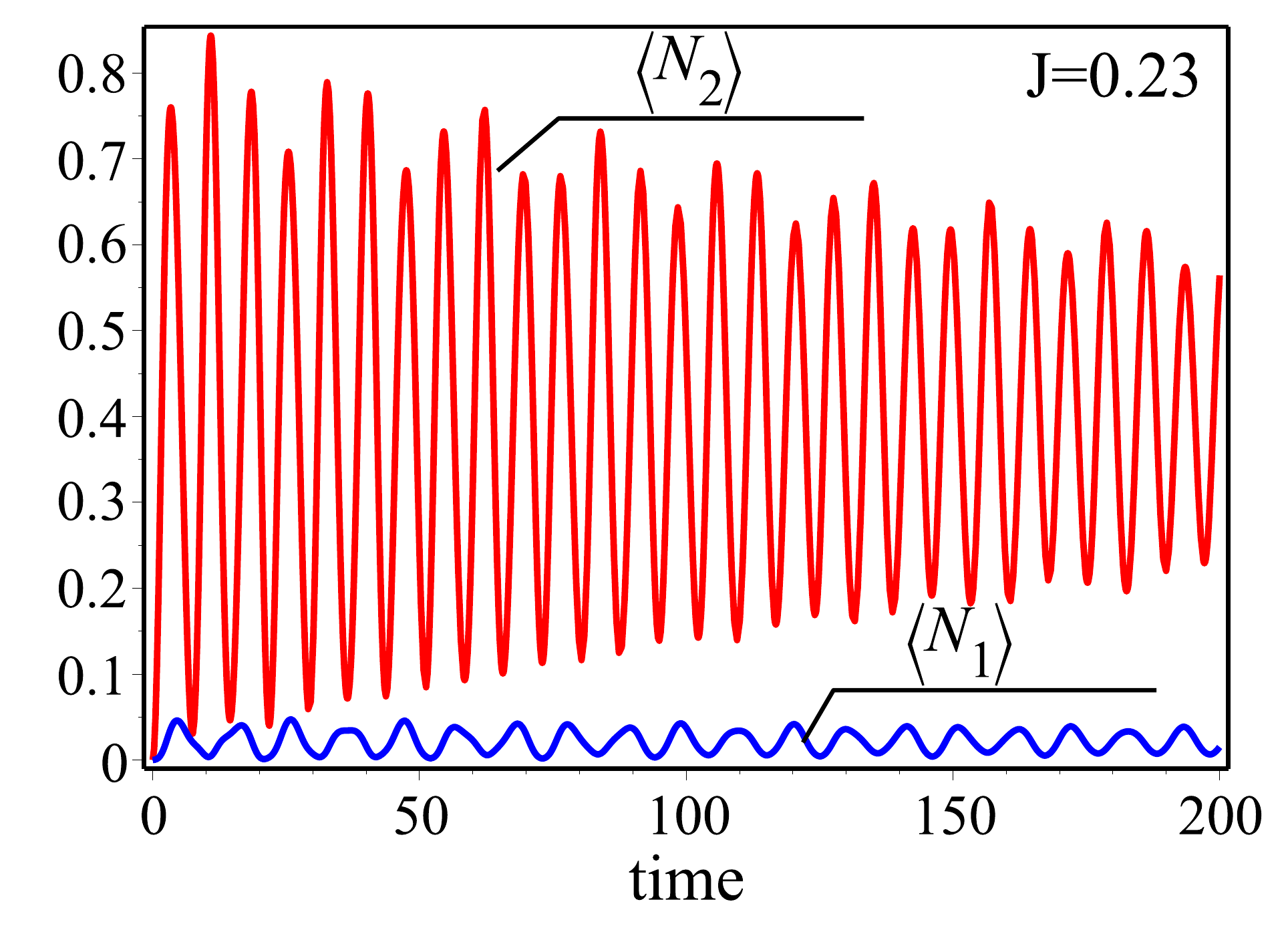}
		\includegraphics[width=5cm, height=4cm]{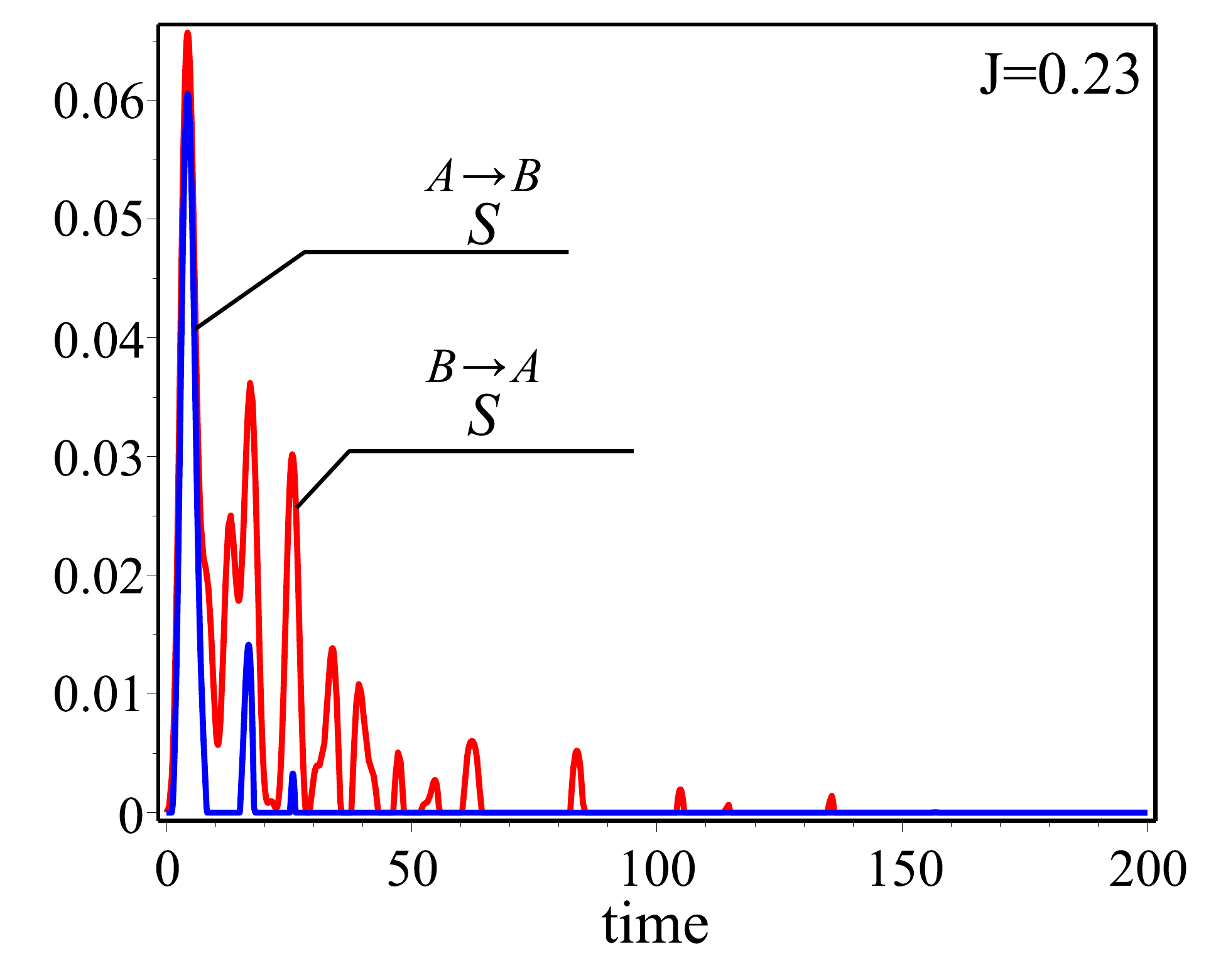}
		\includegraphics[width=5cm, height=4cm]{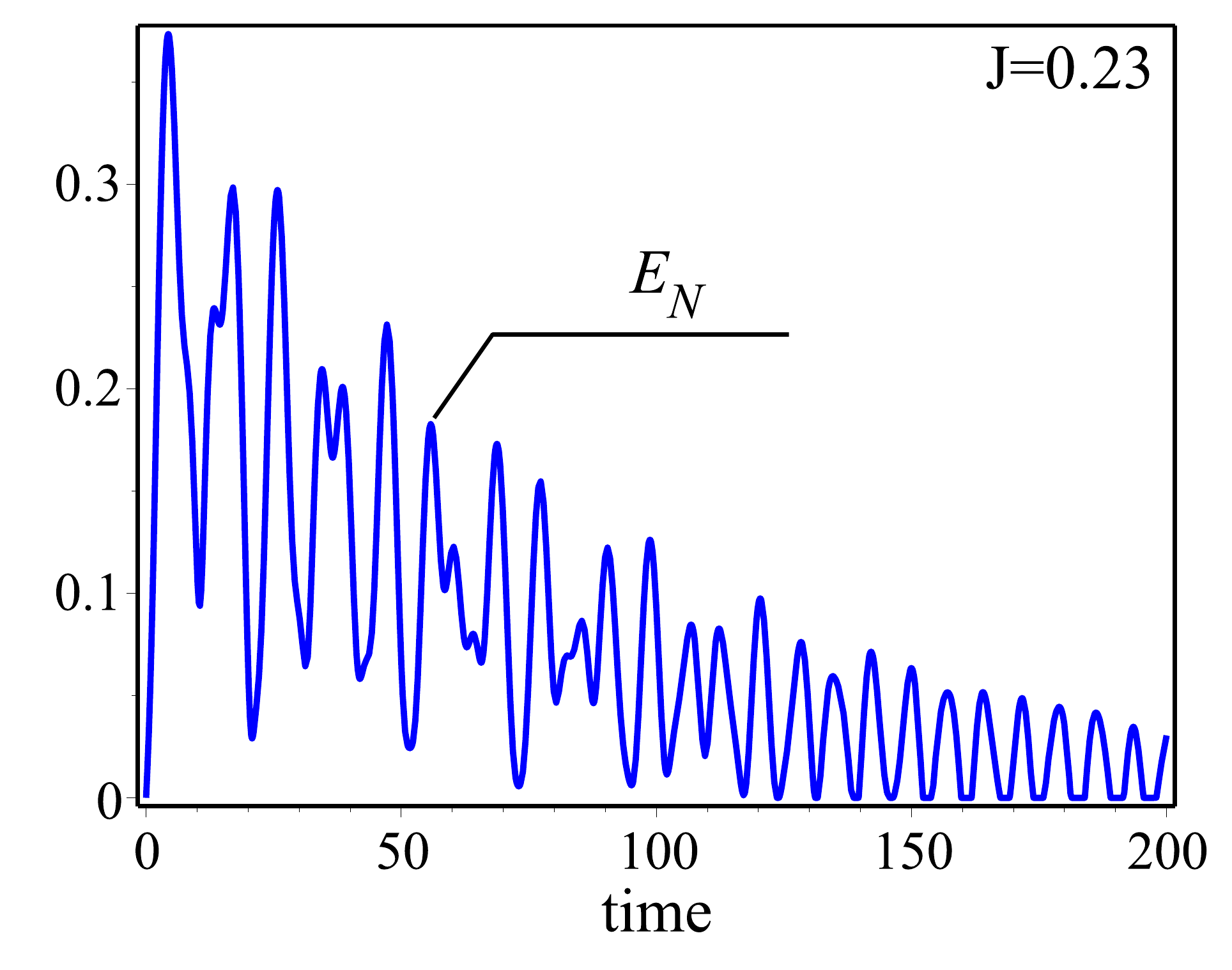}
		\includegraphics[width=5cm, height=4cm]{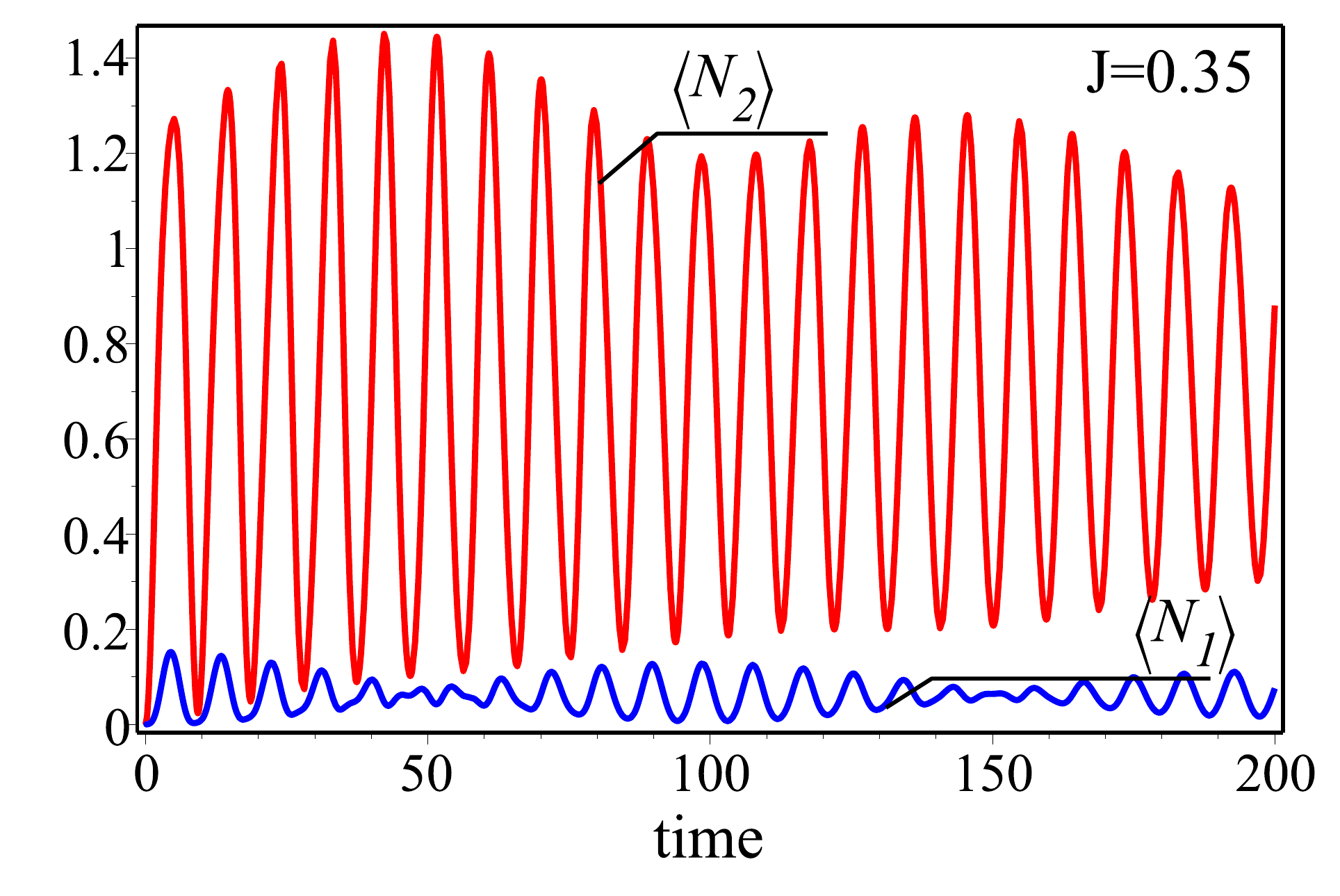}
		\includegraphics[width=5cm, height=4cm]{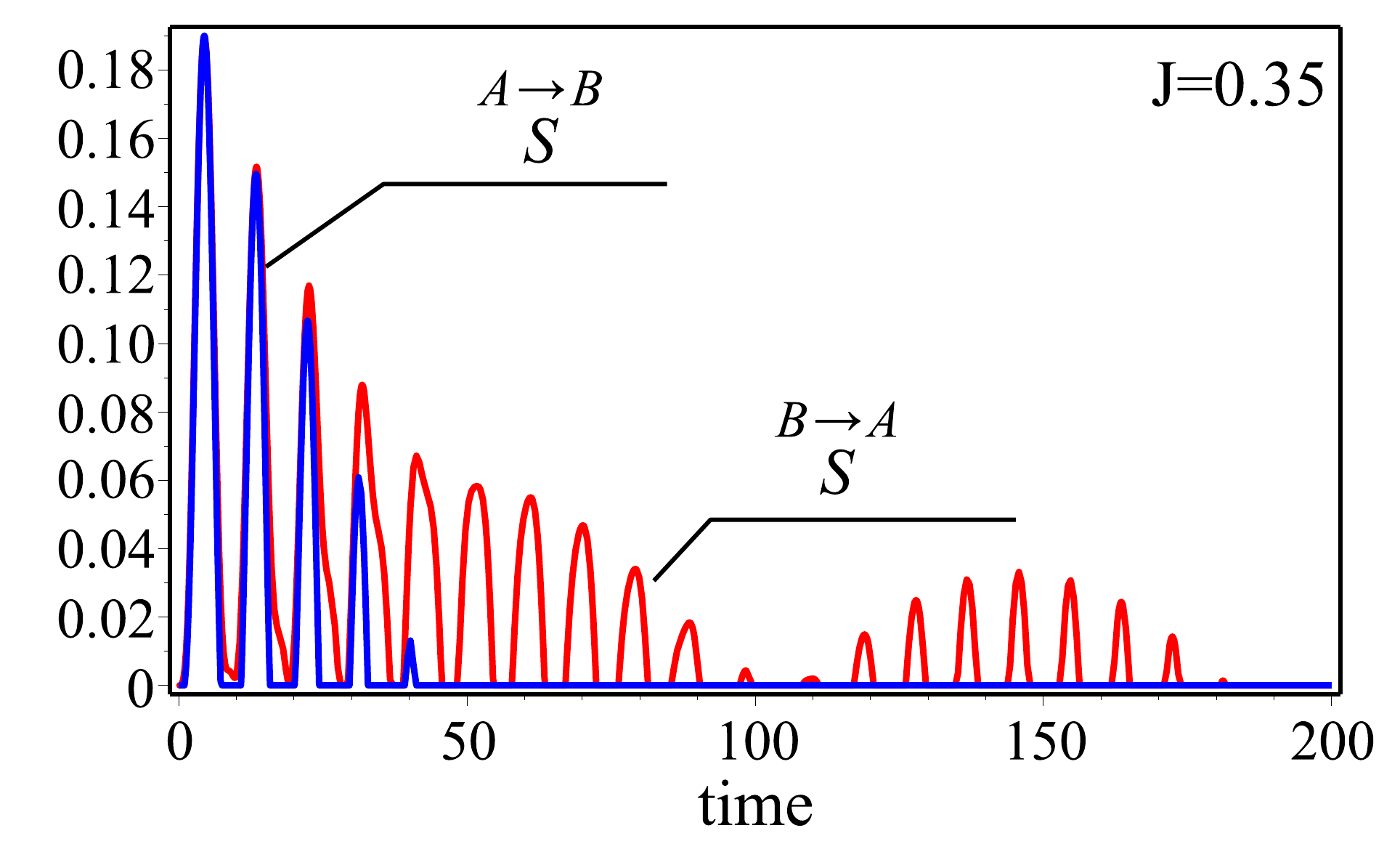}
		\includegraphics[width=5cm, height=4cm]{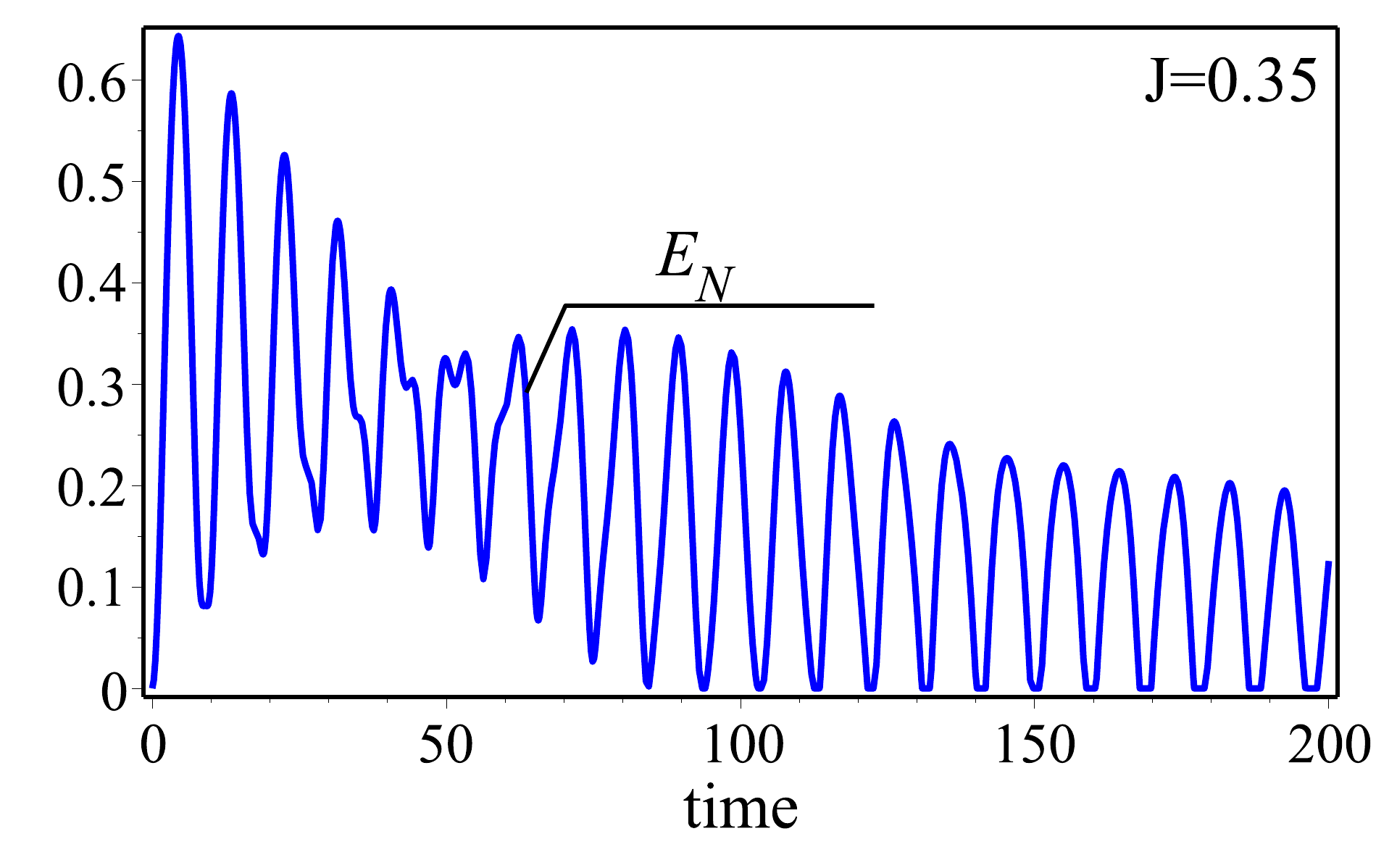}
		\includegraphics[width=5cm, height=4cm]{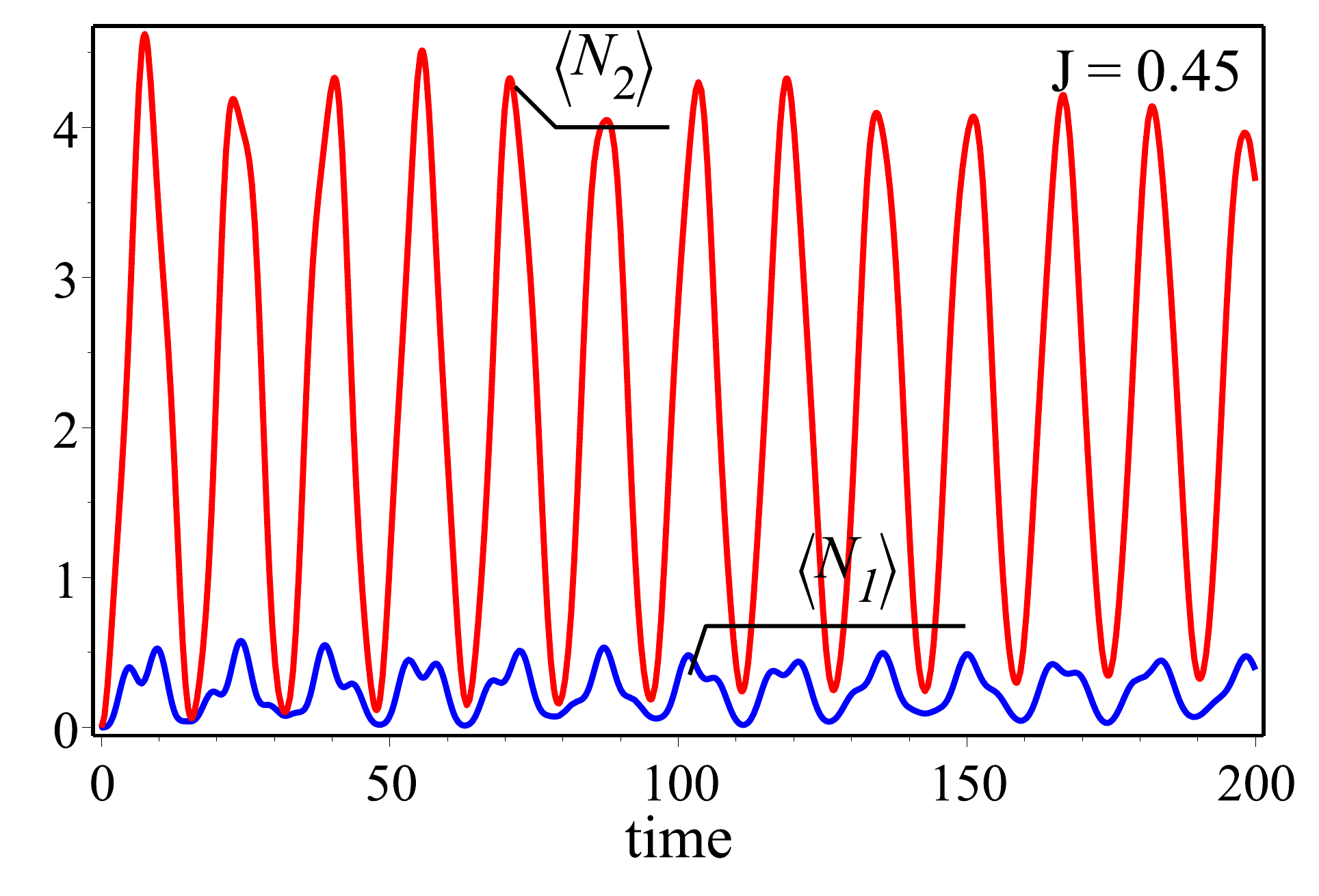}
		\includegraphics[width=5cm, height=4cm]{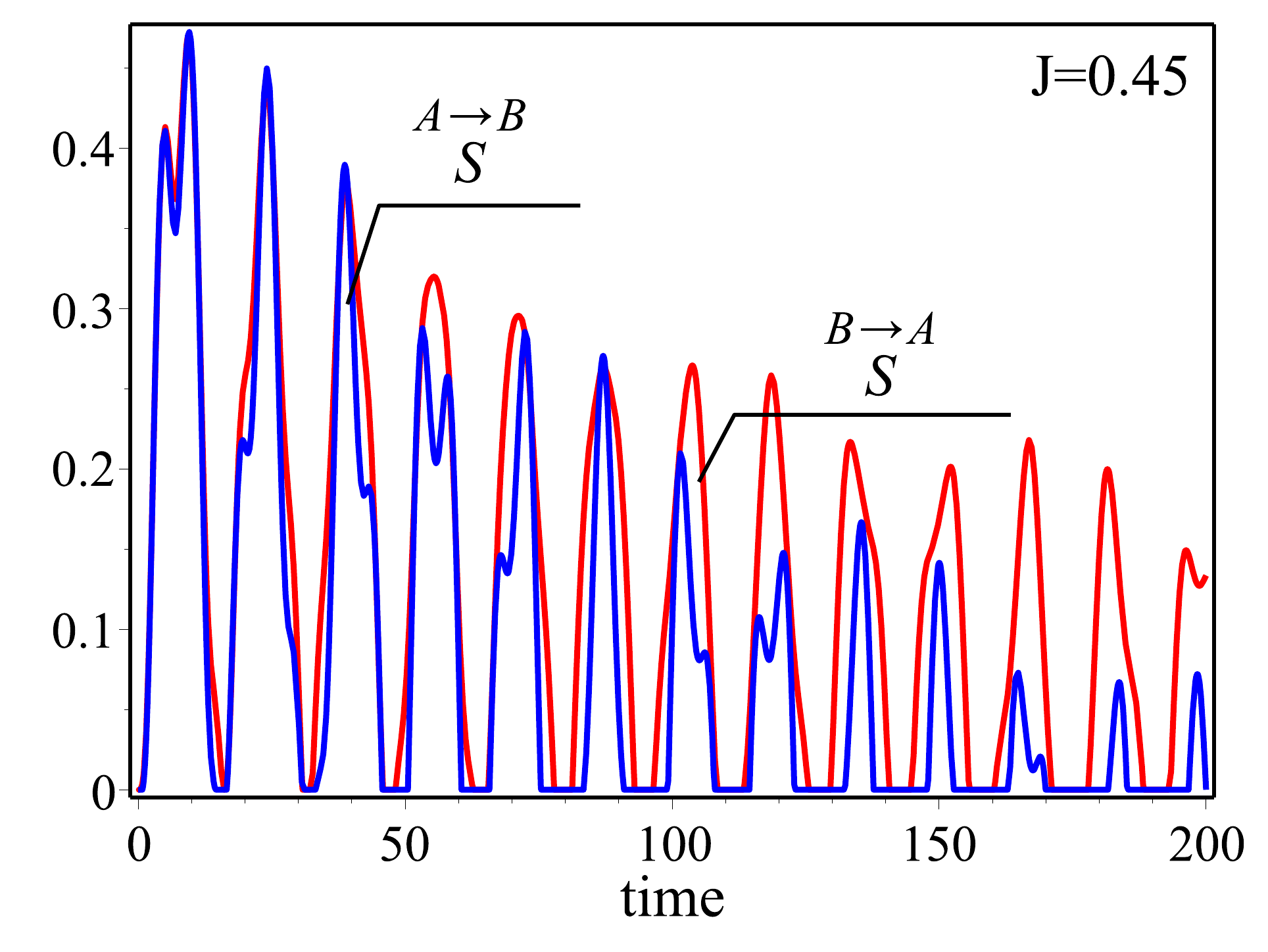}
		\includegraphics[width=5cm, height=4cm]{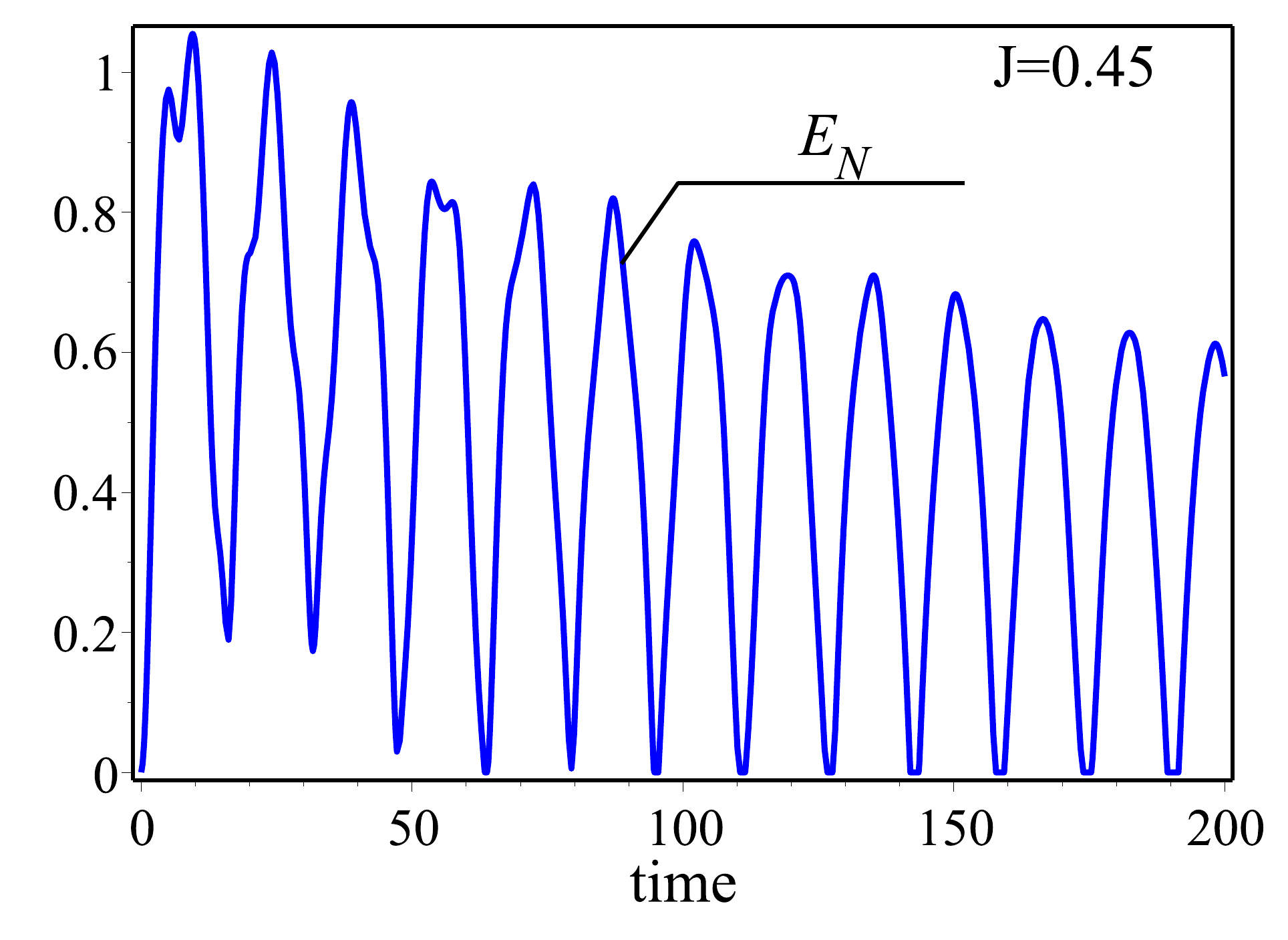}
		\includegraphics[width=5cm, height=4cm]{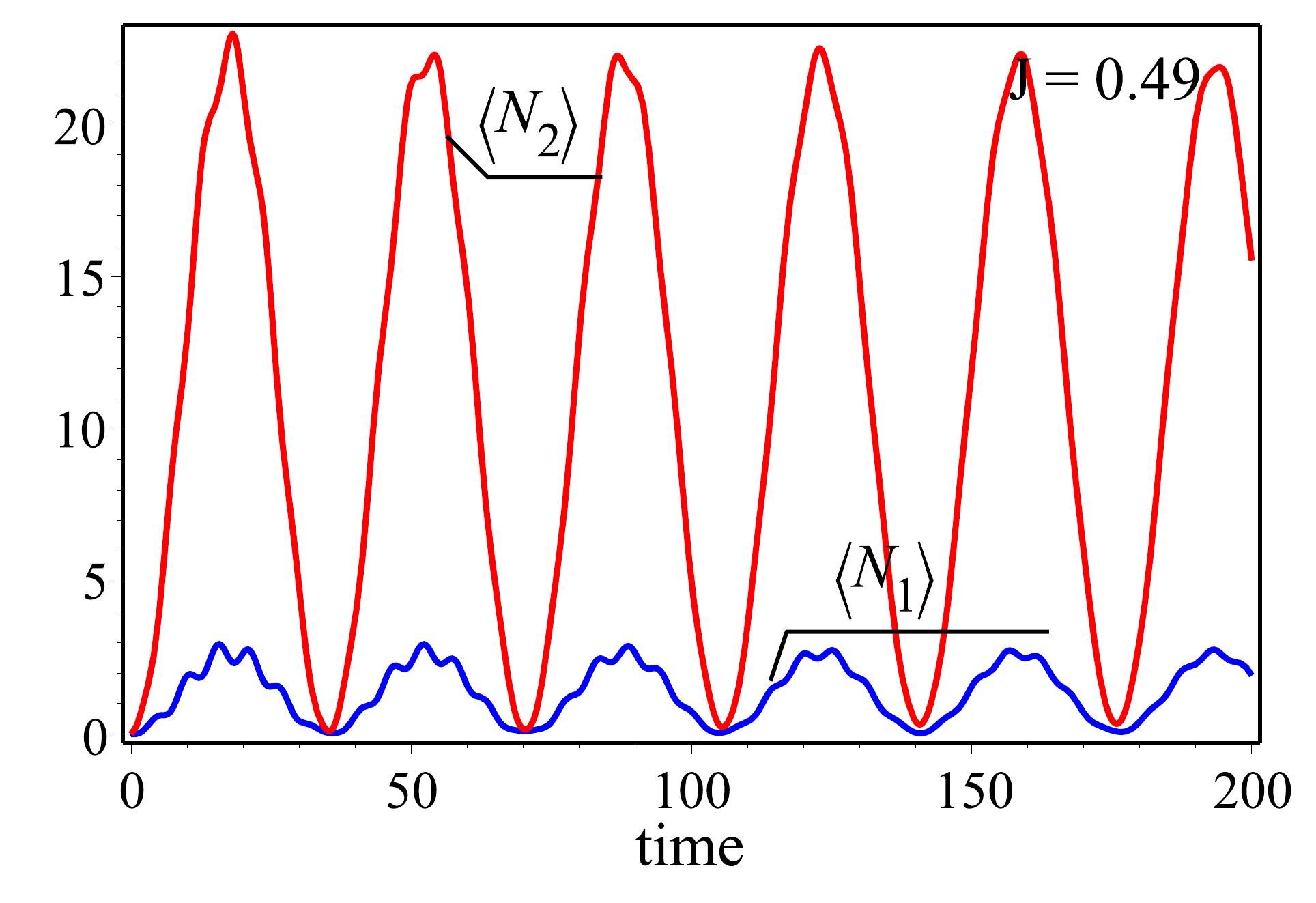}
		\includegraphics[width=5cm, height=4cm]{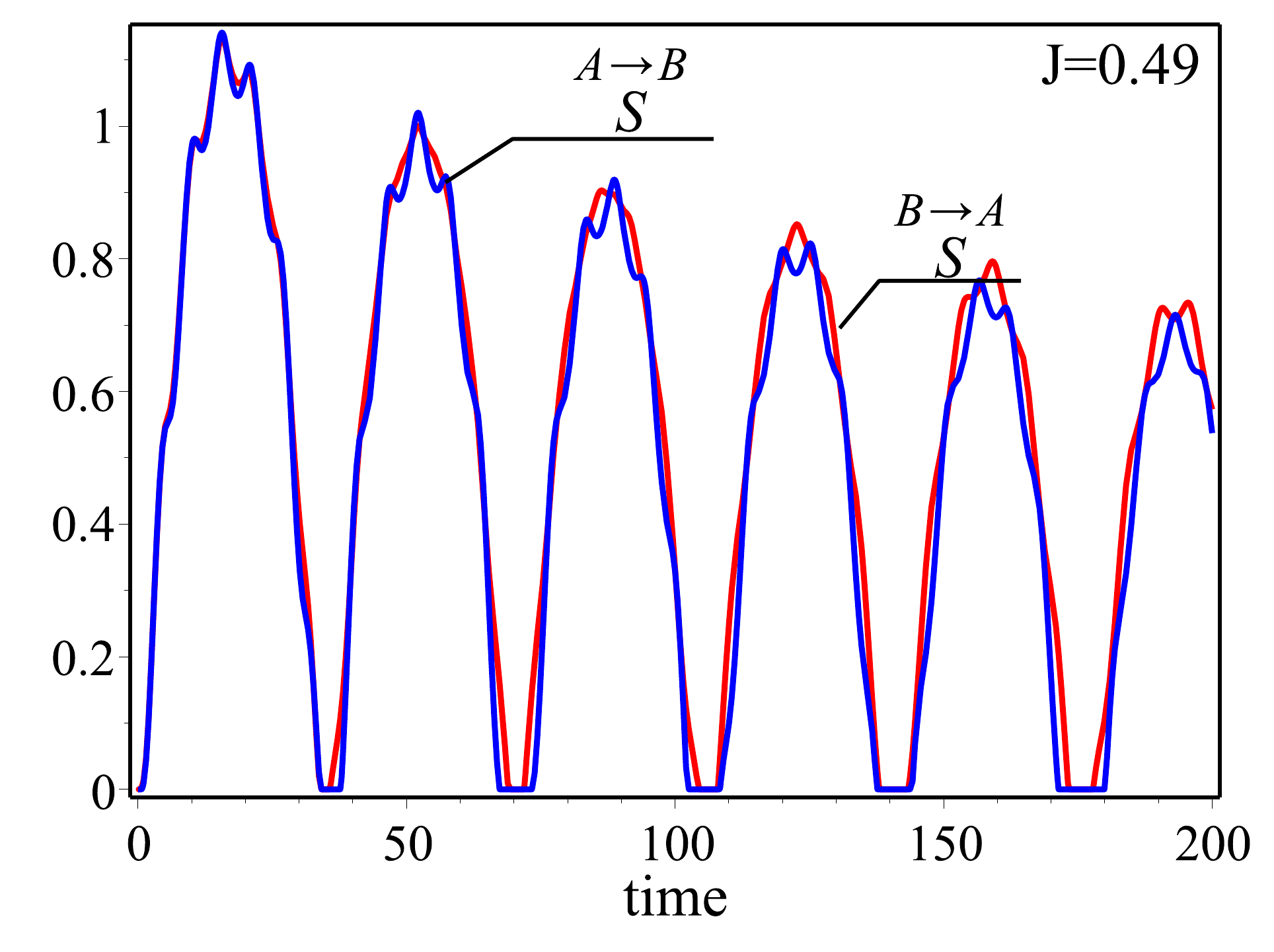}
		\includegraphics[width=5cm, height=4cm]{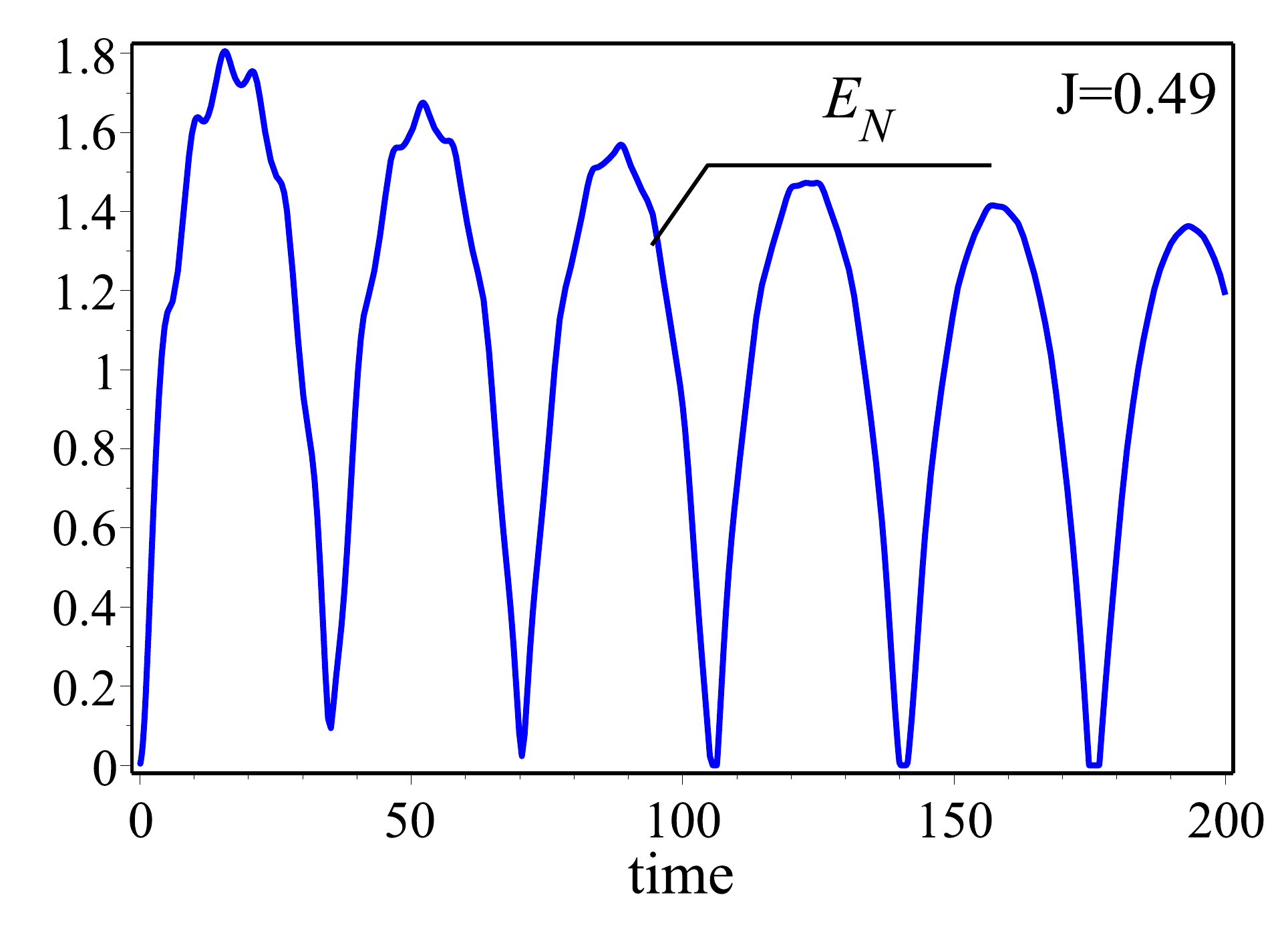}
		\caption{(color online) In the first column, we plot the dynamics of excitations in both oscillators. In the second column, we plot the dynamics of quantum steering for both cases. In the third one, we present the dynamics of entanglement. We  choose $\Gamma=100$, $\omega_1=1$ and $\omega_2=0.5$. }\label{fig3c}
	\end{figure} 

\section{Conclusion}
We have investigated the dynamics of quantum entanglement, quantum steering, and their interconnection   with virtual excitations generated by the counter-rotating terms of the Hamiltonian. The physical system of interest is a system made of two coupled harmonic oscillators coupled via $J\hat{x}_1\hat{x}_2$. 
The system was decoupled and the diagonalisation of the Hamiltonian was obtained after appropriate transformations. 
The initial density matrix is considered to be a separable ground state, namely $|00\rangle \langle 00|$. Subsequently, we have exactly resolved the Milburn master equation to end up with the covariance matrix.

Our findings show that intrinsic decoherence dramatically alters the profile of the oscillatory activity experienced by the three quantities.
We've established that the system has a steady state of entanglement and virtual excitations due to dissipation.
It is shown that steering is less stable than entanglement. Then,
we have demonstrated that coupling ultra-strongly with large anisotropy synchronizes the excitations between the harmonic oscillators. 
As a result, correlations were found to survive for long periods of time in simulations.
Virtual excitations are shown to have a large influence on quantum correlations, which is an issue that can be employed in quantum information science to quantify, regulate, and process quantum resources utilizing virtual excitations.
We have demonstrated the feasibility of preventing inherent decoherence in quantum systems by modifying the experimentally available ultra-strong coupling value.

\end{document}